\documentclass[11pt,twoside]{article}
\usepackage{authblk}
\usepackage{float}
\usepackage{sectsty}
\usepackage{graphicx}
\usepackage{caption}
\usepackage{subcaption}
\usepackage{url} 
\usepackage{multirow}
\usepackage{blindtext}
\usepackage{titlesec}
\usepackage{setspace}   
\usepackage{threeparttable}
\usepackage{amsmath}
\usepackage{rotating}
\usepackage{xcolor}
\usepackage{booktabs}
\usepackage{manyfoot}%
\usepackage{changepage}
\usepackage{threeparttable}
\usepackage{adjustbox}
\usepackage{placeins}
\usepackage{apacite}
\usepackage{natbib}
\usepackage[most]{tcolorbox}
\usepackage{verbatim}

\providecommand{\keywords}[1]
{
  \small	
  \textbf{\textit{Keywords---}} #1
}

\title{The Benefit of Collective Intelligence in Community-Based Content Moderation is Limited by Overt Political Signalling}
\author[a,1]{Gabriela Juncosa}
\author[b, c, 1]{Saeedeh Mohammadi}
\author[d]{Margaret Samahita}
\author[b, c, *]{Taha Yasseri}

\affil[a]{Department of Network and Data Science, Central European University, Austria}
\affil[b]{School of Mathematics and Statistics, University College Dublin, Dublin, Ireland}
\affil[c]{Centre for Sociology of Humans and Machines (SOHAM), Trinity College Dublin and Technological University Dublin, Dublin, Ireland}
\affil[d]{School of Economics, University College Dublin, Dublin, Ireland}
\affil[1]{These authors contributed equally to this work.}
\affil[*]{The corresponding author}
\date{}
\begin{document}
\maketitle	
\doublespacing

\begin{abstract}
Social media platforms face increasing scrutiny over the rapid spread of misinformation. In response, many have adopted community-based content moderation systems, including Community Notes (formerly Birdwatch) on X (formerly Twitter), Community Notes on Meta, and Footnotes on TikTok. However, research shows that the current design of these systems can allow political biases to influence both the development of notes and the rating processes, reducing their overall effectiveness. We hypothesise that enabling users to collaborate on writing notes, rather than relying solely on individually authored notes, can enhance the overall quality of their notes. To test this idea, we conducted an online experiment in which participants jointly authored notes on politically misleading posts. We find that collaboration improves the helpfulness of notes, although the average effect depends on the interactional context. In particular, the benefits of collaboration decline when participants are made aware of one another’s political affiliations. We also find that politically diverse teams improve note quality when evaluating Republican posts, while team composition does not meaningfully affect note quality for Democrat posts. These findings underscore the complexity of community-based content moderation and highlight the importance of understanding group dynamics and political diversity when designing more effective moderation systems.
\end{abstract}

\section*{Introduction}

Misinformation and disinformation permeate all media forms but are particularly amplified on social media platforms due to rapid content dissemination \citep{thai2016big}. An investigation during the 2016 U.S. election found that the average American encountered between one and three posts from known ``fake news'' publishers \citep{allcott2017social}. False information spreads faster and more widely than truthful content, with automated users, such as bots, amplifying election-related posts \citep{vosoughi2018spread}. Moreover, the interaction between individual exposure and confirmation bias exacerbates the impact of misinformation, as people are more likely to believe content that aligns with their political views \citep{frenda2013false, murphy2019false}. Misleading content, defined as factually accurate but lacking context, such as satire, citizen journalism, or one-sided reporting, presents an additional challenge for content moderators. Its prevalence on social media platforms is driven by character limitations and the broadcast-like nature of platforms such as X \citep{wei2017learning, Ecker2014TheHeadlines, Cha2012TheTwitter}. Identifying misleading content is considerably more challenging than identifying misinformation or disinformation, posing a persistent threat to the quality of content online.

To address these challenges, social media platforms have implemented a variety of strategies. Traditional fact-checking relies on human experts to review content and assess its truthfulness. While effective in some cases, this approach is neither scalable nor cost-efficient \citep{hassan2015quest}. Automated algorithms offer scalability and faster processing than expert fact-checking; however, their effectiveness is constrained by the quality and biases of the training data, which can reinforce biases in content moderation \citep{truong2025delayed}. Recent advances in generative AI provide promising solutions to these limitations \citep{burton2024, cui2024,mohammadi2026ai}, but studies indicate that such methods are less effective in the presence of social influence \citep{lu2022effects}.

More recently, platforms have begun experimenting with community-based fact-checking solutions, such as X’s Community Notes \citep{NBCNewsTwitterBirdwatch}, Footnotes on TikTok \citep{presser2025tiktok}, and Facebook's Community Notes initiative \citep{meta2025}. These approaches leverage the benefits of collaborative problem-solving \citep{estelles2012towards, woolley2010evidence, riedl2021quantifying}, enabling users to assess the accuracy of online content collectively. In Community Notes, contributors write contextual notes on posts to clarify or correct information, and other contributors then rate these notes for helpfulness. Notes that receive positive ratings from an ideologically diverse set of raters are ultimately displayed beneath the original post on the platform’s timeline \citep{communityNotes}.  

Preliminary studies suggest that Community Notes is effective in identifying misleading content \citep{saeed2022crowdsourced} and is considerably more scalable and cost-efficient than expert-driven fact-checking \citep{martel2024crowds}. Moreover, when notes appear beneath a post, that post spreads less widely and deeply across the platform \citep{Slaughter2025}.

However, research shows that in the four years since Community Notes’ inception on X, only 13.55\% of posts with at least one proposed note eventually have a note attached to them on the timeline. Among those that do, the process takes an average of 26 hours before a note is displayed \citep{mohammadi2025birdwatch}. Such issues, which undermine the effectiveness of the content moderation process \citep{truong2025delayed}, stem from design limitations in the current Community Notes implementation.

Simulation studies demonstrate that the existing Community Notes framework is highly sensitive to rater bias and in-group preferences, leading to the suppression of many genuinely helpful notes \citep{truong2025community}. Observational analyses of the Community Notes dataset corroborate these findings, showing that partisanship strongly shapes contributors' evaluations: individuals disproportionately flag counter-partisan content as misleading and rate co-partisan notes as more helpful \citep{yasseri2023can, allen2022birds}. 

One proposed explanation is that within Community Notes, the notion of collaboration is not fully implemented: notes are written individually, and contributors rate each other’s notes rather than collaborating directly to develop them, unlike other successful collaborative projects such as Wikipedia \citep{yasseri2023can}. 

Community-based solutions may benefit from more explicitly collaborative approaches to problem-solving. Unlike crowd-sourced systems that rely on aggregating independent contributions, community-based models harness Collective Intelligence (CI)—the ability of a group to solve complex problems and make joint decisions \citep{estelles2012towards, malone2010collective, woolley2010evidence, riedl2021quantifying}. Research shows that CI can enhance political trust and promote well-informed, inclusive political discourse \citep{moore2014democratic, mergel2013implementing}, and the current design of Community Notes and its equivalents in other platforms does not facilitate this.

\begin{figure}
    \centering
    \includegraphics[width=0.9\linewidth]{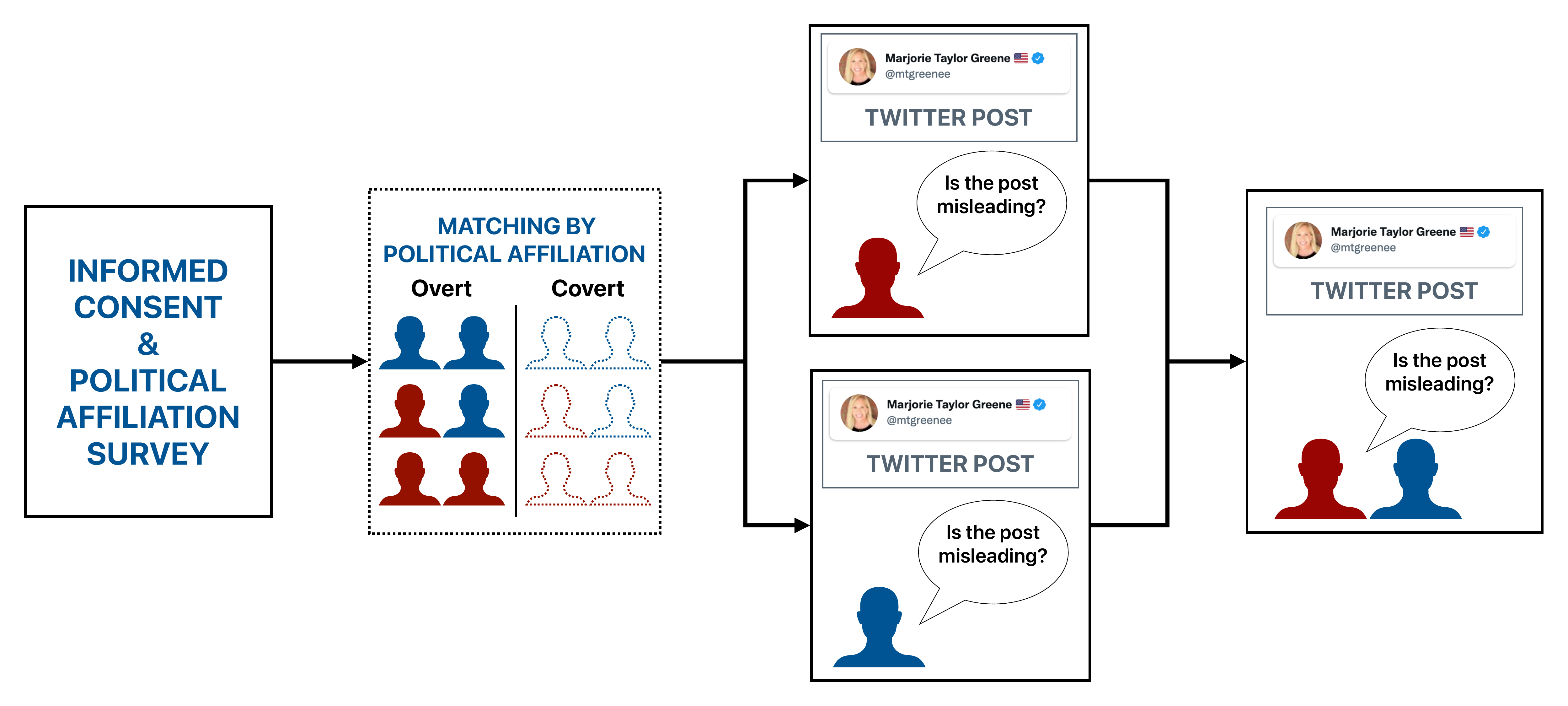}
    \caption{\textbf{Experimental design.} Participants wrote individual and collaborative notes for 40 political posts sourced from Democrat and Republican accounts. We randomly paired participants with partners who either shared or differed in political affiliation, forming three group types: Democrat–Democrat (DD), Republican–Republican (RR), and Democrat–Republican (DR). Each team was then randomly assigned to one of two treatments: the overt condition, in which participants could see each other's political affiliations, or the covert condition, in which affiliations were not disclosed.}
    \label{FlowChart}
\end{figure}

To examine whether a more collaborative community-based design can improve content moderation, we conducted a controlled online experiment with 432 participants recruited from Prolific. We first asked each participant to write an individual note to provide additional context for a social media post. Subsequently, we paired participants into teams and instructed them to communicate via a chat interface to revise and agree on a joint version of their note. We randomly paired the participants with partners sharing the same or opposing political affiliation, resulting in three group configurations: (1) Democrat–Democrat (DD), (2) Republican–Republican (RR), and (3) Democrat–Republican (DR). In addition, we randomly assigned the teams to one of two treatment conditions. In the overt treatment, participants were informed of their partner’s political affiliation, whereas in the covert treatment, this information was not disclosed. Figure~\ref {FlowChart} presents a schematic overview of the experimental design.

The sequential structure of our framework, in which participants first develop their notes independently and subsequently collaborate on a joint version, is consistent with research on effective teamwork. This research suggests that teams benefit when members first formulate and record their ideas independently before sharing and integrating them with their teammates’ ideas \citep{thompson2003improving}. Accordingly, participants in our experiment first considered the post and drafted individual notes, which then served as essential inputs to the subsequent discussion and collaborative revision.

Subsequently, to evaluate the perceived quality of the notes, we recruited 1,610 participants from Prolific who self-identified as Democrats or Republicans and asked them to rate the notes' helpfulness on a scale of 0 to 10. Additionally, we asked a panel of experts to conduct the same evaluation and provide a neutral assessment of note quality.

Our experimental design allows us to explore not only the overall benefits of collaboration but also how political diversity and identity signalling shape the quality of collective output. Through this approach, we aim to advance understanding of how team dynamics influence the effectiveness of community-based content moderation systems.

Our experimental design tests the following hypotheses:

\begin{adjustwidth}{1cm}{1cm}
\textbf{H1.} Teams outperform individuals in writing contextual notes for misleading political content.
\end{adjustwidth}

Although collaboration can improve performance, an important ingredient of collective intelligence is diversity within the group. Teams that are diverse in their perspectives and backgrounds tend to perform better on complex cognitive tasks than homogeneous ones. Previous research demonstrates that diversity can enhance the quality of collaborative output, particularly in domains involving information verification and knowledge production \citep{shi2019wisdom, allen2021scaling, pennycook2019fighting, arazy2011information, yasseri2012dynamics}. \citet{shi2019wisdom} find, for example, that higher political polarisation among Wikipedia editors correlates with higher quality articles, especially on political topics, because conflict forces contributors to strengthen their reasoning and engage with opposing perspectives. 

In Community Notes, diversity is currently leveraged only at the rating stage. Our experiments allow us to examine whether introducing political diversity at the writing stage further enhances performance. We hypothesise:

\begin{adjustwidth}{1cm}{1cm}
\textbf{H2.} Politically diverse groups outperform politically aligned groups in writing contextual notes for misleading political content.
\end{adjustwidth}

Diversity can also introduce challenges, such as miscommunication \citep{straub2023cost, weber2003cultural, cannon1993shared, cronin2007representational}, reduced trust \citep{alesina2002trusts, putnam2007diversity}, and potential interpersonal conflict \citep{cui2024}. In particular, political partisanship can generate friction within collaborative tasks, as studies show that individuals often hold deep negative feelings toward members of the opposing party \citep{iyengar2012affect, abramowitz2016rise}. These negative attitudes can be further intensified through identity signalling. Identity signalling refers to any form of communication that conveys one’s membership in a social or political group \citep{van2022strategic}. Political identity signalling can exert strong effects on individuals’ judgments and behaviours \citep{iyengar2015fear}. Such signals can range from overt expressions of partisanship to more subtle cues, such as taking a stance on a political issue. Prior research indicates that the more explicit these signals are, the less likely cross-partisan interactions will lead to consensus or depolarisation \citep{guilbeault2018social}. Therefore, we further hypothesise:

\begin{adjustwidth}{1cm}{1cm}
\textbf{H3.} When political identity is overtly known, teams produce less helpful contextual notes for misleading political content.
\end{adjustwidth}
\section*{Results}
In our initial hypothesis, we proposed that collaboratively written notes would be rated as more helpful than those authored individually. To test this, we compared the average helpfulness scores of notes written by individuals before collaboration with notes written collaboratively within their teams.

\begin{figure}[!htbp]
\centering
\includegraphics[width=0.55\linewidth]{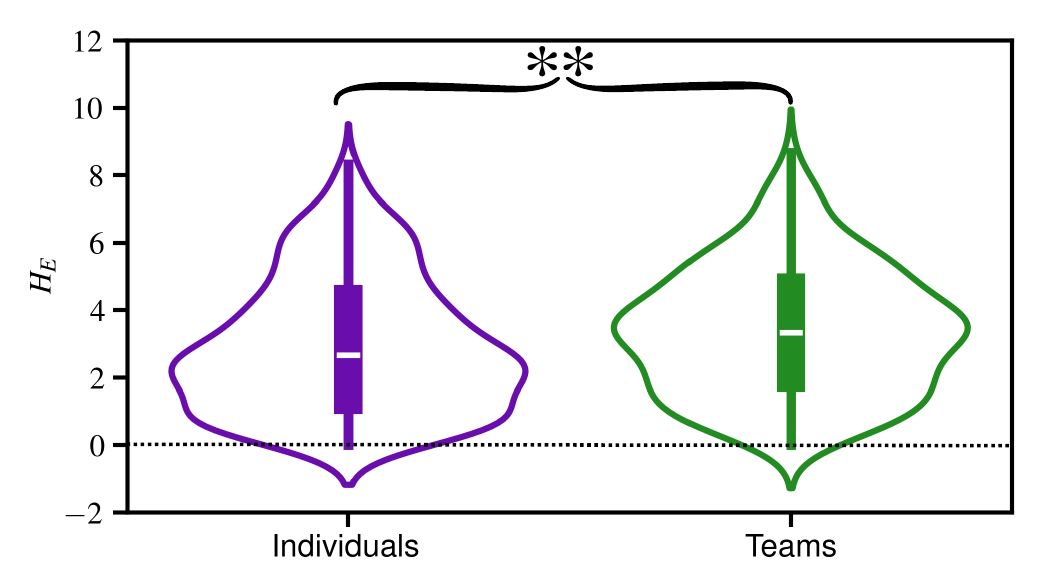}
\caption{\textbf{Distribution of helpfulness scores according to experts.} The distribution of helpfulness scores according to expert ratings, $H_{\rm E}$, for individually written notes and collaboratively written notes. ** indicates $p < 0.01$.}
\label{collaboration_figure}
\end{figure}

Figure~\ref{collaboration_figure} depicts the distribution of helpfulness scores based on expert ratings ($H_{\rm E}$), and Table~S1 reports the results of two-sample hypothesis tests comparing individually authored notes with collaboratively authored ones. Both expert and crowdsourced ratings are reported.

Expert ratings reveal that collaboratively written notes are significantly more helpful than individually written notes (individuals: $M = 2.9$; teams: $M = 3.4$; $t(430) = -2.8$; $p < 0.01$; Cohen’s $d = 0.27$).  The average relative difference between individually and collaboratively written notes is approximately 17\%. While the difference observed in the crowdsourced ratings is smaller than that found in the expert evaluations, it remains statistically significant among Republican raters, though not among Democrat raters. Nevertheless, the average helpfulness of collaboratively written notes increases across all evaluation groups. These results suggest that collaboration can effectively enhance the perceived quality of notes. However, this must be interpreted in light of the fixed task order. Because participants always completed the individual task before the collaborative stage, the observed improvement cannot be attributed exclusively to collaboration and may partly reflect repetition and additional reflection. Future research could address this limitation.

As a robustness check, we examined whether the helpfulness scores contained distinct helpful and unhelpful modes, which could indicate that some participants put less effort into the task than others. We therefore calculated the Bimodality Coefficient (BC) separately for individual and team notes. Based on expert ratings, the coefficients were $\rm{BC}_{\rm{Individuals}} = 0.429$ and $\rm{BC}_{\rm{Teams}} = 0.432$, providing no evidence of bimodality in either distribution. The same pattern emerged in ratings provided by Democrats ($\rm{BC}_{\rm{Individuals}} = 0.396$; $\rm{BC}_{\rm{Teams}} = 0.365$) and Republicans ($\rm{BC}_{\rm{Individuals}} = 0.349$; $\rm{BC}_{\rm{Teams}} = 0.363$). Overall, these results suggest that the observed mean differences were not driven by a distinct subgroup producing exceptionally unhelpful notes.

\begin{figure}[!htbp]
\centering
\includegraphics[width=.9\textwidth]{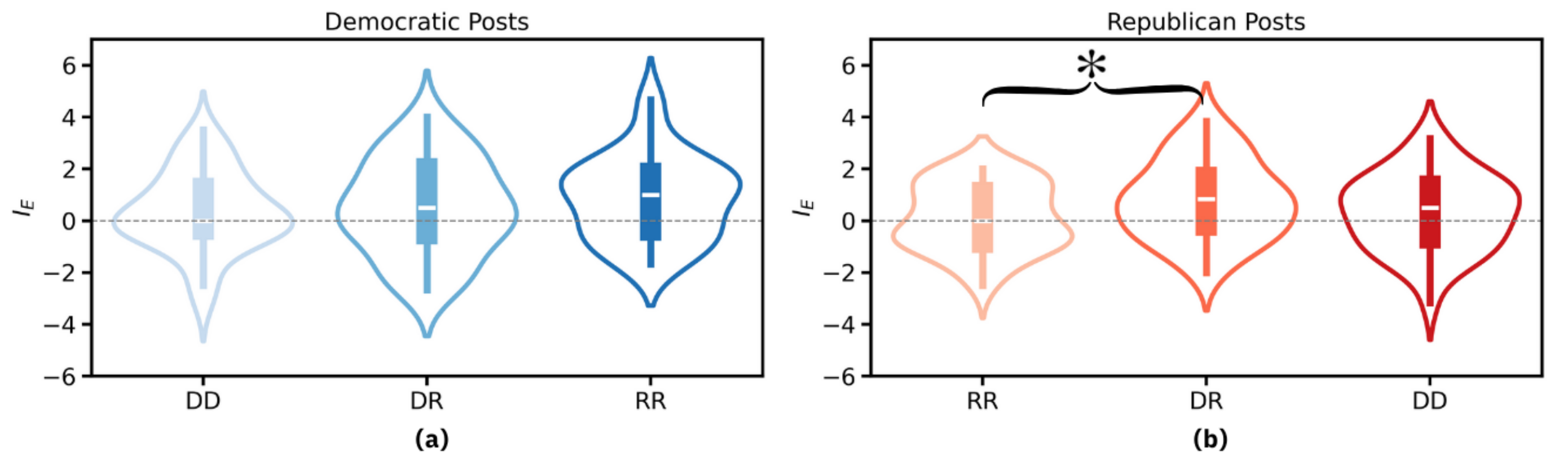}
\caption{\textbf{Distribution of ${I}_{\rm{E}}$ of notes categorised by group's political composition} (a) Distributions of ${I}_{\rm{E}}$ of Democrat posts across groups with different political compositions: Democrat teams (DD), diverse teams (DR), and Republican teams (RR). (b) Distributions of ${I}_{\rm{E}}$ of  Republican posts across groups with different political compositions: Republican teams (RR), diverse teams (DR), and Democrat teams (DD). * indicates $p < 0.05$.}\label{political_composition_figure}
\end{figure}

To better understand the dynamics in which collaboration can lead to higher--quality outcomes, we investigate group dynamics in greater detail, including political partisanship and political identity signalling. First, we examine the role of political partisanship in this process. We further analysed notes based on the political composition of the pair who authored them and the political leaning of the social media post. Figure~\ref{political_composition_figure} presents the distribution of the change in helpfulness of the notes evaluated by experts (${I}_{\rm{E}}$) across different team types, and Table~S2 reports the results of two-sample hypothesis tests comparing their performance. For Republican posts, the reference category is the Republican team (RR), with the diverse (DR) and the Democrat (DD) teams as treatment groups. For Democrat posts, the reference category is the Democrat team (DD), with the diverse (DR) and the Republican (RR) teams as treatment groups.

For Democrat posts, neither RR nor DR teams outperform the DD team, indicating that performance is comparable across all team compositions. In contrast, for Republican posts, the DD teams perform similarly to the RR teams, while the politically diverse DR teams produce significantly better notes than the RR teams (RR: $M = 0.11$; DR: $M = 0.78$; $t(70) = -2.0$; $p < 0.05$; Cohen’s $d = 0.48$). These findings suggest that introducing diversity within teams enhances performance only for Republican posts (See Tables~S4 and S5 for more details).

To examine whether this pattern is driven by the number of team members politically aligned with the post, we introduce a measure of post alignment, which captures the number of participants in the team whose political affiliation aligns with the post (details are provided in Methods). We then ran linear regressions with ${I}_{\rm{E}}$, ${I}_{\rm{D}}$, and ${I}_{\rm{R}}$ as outcome variables. The results (Table~S8) show that post alignment is not a significant predictor of changes in helpfulness. This suggests that merely introducing an out-group member into the note-writing process does not necessarily improve note quality.

To examine the effects of political identity signalling, we introduced an additional treatment condition by informing some teams of their partner’s political identity while withholding this information from others. Accordingly, we partitioned the results into two treatment groups: those in which political identity was explicitly disclosed (overt condition) and those in which it remained undisclosed (covert condition).

As a manipulation check, we additionally coded the chat messages using an LLM-assisted qualitative coding procedure described in the Supplementary Information. Specifically, we measured partisan content salience, partisan identity salience, disagreement, and concessionary behaviour in each chat. Although the differences between overt and covert conditions are not statistically significant across these chat dimensions, the descriptive patterns indicate higher average partisan identity salience, disagreement, and concessionary behaviour in the overt condition (Figure~S16).

\begin{figure}[!htbp]
\centering
\includegraphics[width = 0.75 \linewidth]{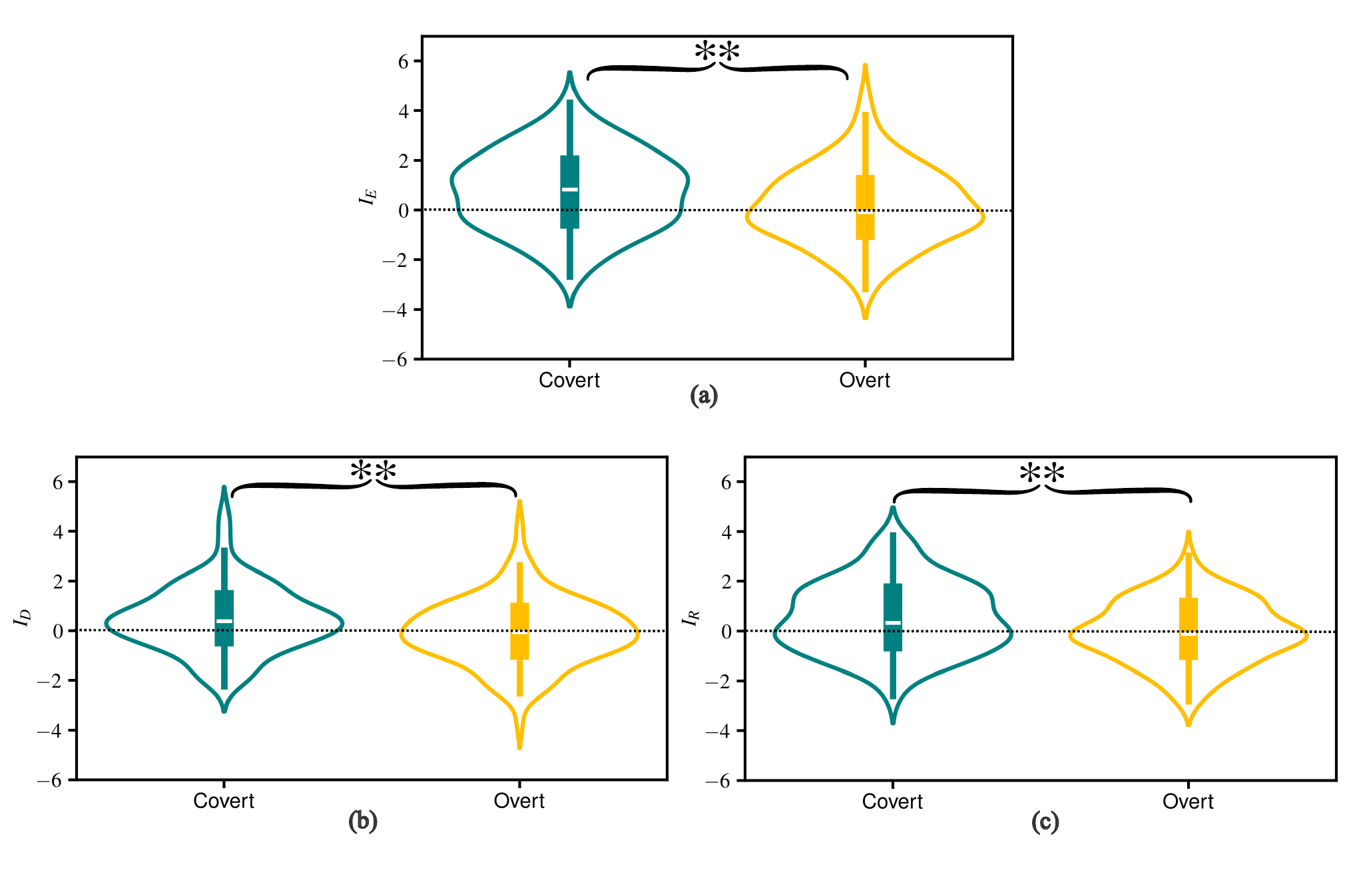}
\caption{\textbf{Distribution of the change in helpfulness for collaborative notes under covert and overt treatments, based on both crowd-sourced and expert assessments.} (a) The distribution of ${I}_{\rm{E}}$, (b) The distribution of $I_{\rm{D}}$, (c) The distribution of $I_{\rm{R}}$ for covert and overt treatments. ** indicates $p < 0.01$}
\label{overt_covert_figure}
\end{figure}

Figure~\ref{overt_covert_figure} illustrates the distributions of ${I}_{\rm{E}}$ alongside the change in the helpfulness of notes according to crowds of Democrats (${I}_{\rm{D}}$) and Republicans (${I}_{\rm{R}}$) across covert and overt treatments. Table~S3 summarises the results of parametric and non-parametric two-sample hypothesis tests comparing the change in helpfulness between covert and overt treatments. 

Expert evaluations indicate that the collaborative advantage in note helpfulness diminishes when political identity is overtly disclosed (Covert: $M = 0.76$; Overt: $M = 0.16$; $t(214) = 2.9$; $p < 0.01$; Cohen’s $d = 0.39$).  This pattern is consistent across Democrat ratings (Covert: $M = 0.45$; Overt: $M = -0.045$; $t(214) = 2.8$; $p < 0.01$, Cohen's $d = 0.38$) and Republican ratings (Covert: $M = 0.54$; Overt: $M = -0.037$; $t(214) = 3.2$; $p < 0.01$, Cohen's $d = 0.44$). Observations across some teams reveal that overt signalling of political identity yields a slightly negative ${I}_{\rm{E}}$, suggesting that collaboratively written notes with overt political signalling are, in some cases, less helpful than those written individually by the same participants. Table~S6 compares the covert and overt conditions across different team compositions.

To examine this pattern more directly, we estimated baseline linear regression models predicting changes in note helpfulness from political identity signalling. Across expert, Democratic, and Republican ratings, the overt condition was associated with significantly smaller improvements in helpfulness (Table~S10), consistent with the comparisons above.

We next examined whether political identity signalling was associated with changes in note characteristics. Separate linear regression models showed that the mean increase in note length ($I_{\mathrm{Length}}$) was significantly smaller in the overt condition than in the covert condition, indicating that teams in the overt condition were less likely to write longer revised notes. In contrast, the difference between conditions in the change in the number of links ($I_{\mathrm{Links}}$) was not statistically significant (Table~S9).

To assess whether these changes in note characteristics could account for the relationship between political identity signalling and helpfulness improvement, we estimated additional regression models controlling for changes in note length and number of links. Both $I_{\mathrm{Length}}$ and $I_{\mathrm{Links}}$ were positively and significantly associated with improvements in helpfulness across all three evaluator groups. However, political identity signalling remained statistically significant after including these controls (Table~S11). Figure~S17 presents the path diagram corresponding to these models.

Finally, we examined whether this pattern differed by team diversity and found that the effect of political identity signalling persists for both politically diverse and politically aligned teams (Table~S12). This suggests that the reduction in collaborative advantage under overt political signalling is not attributable solely to differences in note length or the number of links, nor is it limited to one type of team composition.

\section*{Discussion}
On January 7, 2025, Mark Zuckerberg announced that Meta would phase out other content moderation strategies and instead rely exclusively on community-based approaches, similar to X’s Community Notes \citep{meta2025}. This announcement signalled a broader shift across social media platforms: shortly thereafter, TikTok introduced “Footnotes”, a similar system for community--based content moderation \citep{presser2025tiktok}.

This transition raises concerns because prior research shows that Community Notes often fail to provide notes promptly, frequently appearing only after the content has already spread widely \citep{mohammadi2025birdwatch, chuai2026community, truong2025delayed}. Both simulation and observational studies indicate that these delays stem from political polarisation among users and in-group preferences \citep{truong2025community, allen2022birds, yasseri2023can}. Scholars argue that the current structure of Community Notes does not fully leverage contributors’ political diversity during the note-writing stage, and that collaboration among contributors during this stage could improve both the efficiency and quality of such systems \citep{yasseri2023can}.

In this study, we tested this idea through an online experiment. We recruited participants and asked them to write a note that provides context for a misleading post, similar to how contributors write notes in Community Notes. Afterwards, we paired participants and asked them to revise their notes together based on their shared judgment. The results supported our first hypothesis: expert evaluations rated collaboratively written notes as more helpful than individually written notes. On average, the helpfulness of notes increased by approximately 17\% after collaboration. The corresponding improvement is smaller when evaluated by crowd raters.

The small size of this effect is substantively important. It suggests that collaboration can improve note quality, but that collaboration alone is not a sufficient design solution. Rather, the benefits of collaboration appear to depend on the interactional environment in which collaboration occurs. This motivated our second hypothesis, which examined whether the partisan composition of teams shapes the extent to which notes improve.

Political diversity is an important foundation of community-based initiatives, which draw on contributors with different perspectives to improve collective outcomes. In community-based content moderation, aggregating judgments from politically diverse contributors can allow individual errors to offset one another as participation increases \citep{martel2024crowds}. Wikipedia provides a prominent example of the value of contributor diversity in collaborative knowledge production. Its success has been attributed both to the diversity of its editors and to the platform’s policies and guidelines \citep{arazy2011information, yasseri2012dynamics}. For instance, studies show that the association between crowd size and article quality is stronger when editors are more diverse \citep{robert2015crowd}. Elsewhere, politically diverse groups of editors have been associated with higher-quality articles than politically homogeneous groups \citep{shi2019wisdom}.

Our research complements these studies by examining whether the benefits of political diversity extend to collaboration in a different and more contested setting: content moderation. We compared how much notes improved across groups with different partisan compositions. The results provide mixed support for this hypothesis: politically diverse teams showed greater improvement when evaluating Republican posts, whereas team composition did not meaningfully affect improvement for Democratic posts.

This finding points to an asymmetry in how Democrats and Republicans engage on social media. Previous research shows that Republicans are more likely to be in an echo chamber \citep{barbera2015tweeting} and that their posts are more frequently flagged as misinformation \citep{renault2025republicans}. While our data do not allow us to identify the underlying cause of this asymmetry, the results suggest that diversity does not automatically enhance collective intelligence in community-based content moderation.

Our final manipulation examined the effect of political identity signalling on notes. In the current Community Notes structure, all contributors use aliases, making it impossible to link contributors to their X accounts \citep{XCommunityNotesAliases}. However, it is still possible to infer a contributor's political identity through their previous notes. To test the effect of political signalling, we compared teams that were explicitly shown their partner’s political identity (overt condition) with teams that were not provided this information (covert condition). The results indicate that when participants were aware of each other’s political partisanship, the improvement in perceived note quality from collaboration was significantly reduced. Although the effect sizes were modest, this pattern was consistent across expert ratings and ratings from crowds of Democrats and Republicans.

Several mechanisms may help explain why making political identity explicit reduced the benefits of collaboration. We first examined whether it could be attributed to observable changes in note characteristics (Table~S9). Participants in the covert condition tended to increase the length of their revised notes more than those in the overt condition. However, the association between political identity signalling and note improvement remained significant after controlling for changes in note length and number of links (Figure~S17), suggesting that these characteristics do not fully account for the observed effect.

Second, we examined whether making political identity visible changed how participants interacted with each other. The manipulation check based on chat messages revealed no statistically significant differences between the overt and covert conditions on the interactional dimensions we examined. Descriptively, however, conversations in the overt condition showed higher average levels of partisan identity salience, disagreement, and concessionary behaviour (Figure~S16). These patterns help contextualise the decline in note quality under overt political signalling. Rather than simply changing the information available to participants, making political affiliation visible may have altered the social context in which collaboration occurred. This interpretation aligns with previous research showing that overt political identity cues can create friction within collaborations and undermine the benefits of collective intelligence \citep{iyengar2012affect, guilbeault2018social}. When political identity is made explicit, participants may become more attentive to partisan considerations. They may evaluate their partner’s suggestions through the lens of political identity rather than solely based on the evidence they provide. In our study, overt partisan signalling is associated with a decline in note quality, suggesting that political identity cues can reduce the positive effects of collaboration and, in some cases, may entirely reverse them, producing outcomes that are worse than if participants had worked individually. 

A key limitation of our study is that our design used a fixed task order: participants always completed the individual note-writing task before entering the collaborative revision stage. This sequencing reflects a common workflow in group problem-solving, where individual reflection often precedes collective discussion, and our experiment formalised this structure. This structure is consistent with research on team creativity, which suggests that collaboration can be more effective when individuals first develop their own ideas before sharing and integrating them with those of their teammates \citep{thompson2003improving}. Nevertheless, the design does not allow us to fully separate the effect of collaboration from possible order or repetition effects. Participants may have improved simply because they had more time to reflect on the post, because they encountered the task a second time, or because the collaborative stage allowed them to refine an already-formed judgment. Accordingly, the finding that collaboration improves note quality should be interpreted in light of this limitation, as the observed improvement cannot be attributed exclusively to collaboration. Future work should therefore use counterbalanced designs that vary the task sequence, or include conditions in which participants complete only the collaborative task, to estimate the independent contribution of collaboration more directly. Taken together, however, our findings fit within a broader literature on the benefits of collaboration in complex problem-solving \citep{woolley2010evidence, estelles2012towards, malone2010collective} and suggest that community-based approaches could enhance the quality of information on social media platforms, under specific conditions.

A second limitation concerns the generalisability of the experimental setting to real-world community moderation systems. Participants in our study were incentivised to communicate and write notes, whereas Community Notes contributors are volunteers. The interaction environment also differed from real-world note writing: participants collaborated in a structured experimental interface, under time and task constraints, rather than in the more open-ended and asynchronous environment of an actual platform. Most importantly, our political identity manipulation was deliberately direct and abstract: participants were explicitly informed of their partner’s political affiliation. In real-world settings, political identity is less likely to be disclosed in such a clear and decontextualised way. Instead, users may infer political orientation indirectly from writing style, note content, account history, or patterns of agreement and disagreement. For this reason, the overt-signalling treatment should be interpreted as a stylised test of the mechanism rather than a direct simulation of everyday platform interaction. Even so, the experiment remains informative because it isolates how making political identity salient can alter collaborative outcomes. This mechanism may also operate in subtler forms in real-world community-based moderation.

A further limitation concerns the possibility that participants may have used external generative AI tools to assist in writing their notes. We have good reason to believe this was unlikely. The experiment was conducted in the summer of 2023, when such tools were not yet widely accessible to most users, particularly for rapid, task-specific assistance of the kind required in our study. This context reduces the likelihood of using external models and strengthens the reliability of our results. Nonetheless, as these technologies continue to evolve, future work should incorporate explicit checks or design features to monitor and account for potential AI-assisted contributions.

Additionally, our study examined duos, the simplest form of a team. Although this design limits our ability to capture more complex group dynamics, it provides a useful foundation for future research on collaboration in larger groups. Future studies should examine larger teams with varying political compositions. This would allow researchers to investigate how the degree and distribution of political diversity influence participation, collaboration, and note quality, and whether the patterns observed here extend to larger groups.

The insights from this study have direct implications for the design and implementation of community-based fact-checking systems on social media platforms. Our results suggest that collaboration can improve the quality of contextual notes, but that platforms should be cautious about making contributors’ political identities visible during the note-writing process. A promising design principle is therefore to encourage exposure to diverse political perspectives without exposing contributors’ political identities. Platforms could, for example, algorithmically match contributors with different inferred perspectives while preserving alias-based anonymity, or create collaborative drafting interfaces in which contributors are exposed to diverse arguments without being shown their partisan identity. Such designs would retain the epistemic benefits of collaborative input while reducing the risk that identity cues trigger stereotyping, distrust, or disagreement. 

Beyond these practical implications, our study sheds light on the dynamics of collective intelligence in the politically sensitive domain of content moderation. It suggests that collaboration and diversity alone may not be sufficient to improve collective outcomes. Rather, the effectiveness of community-based moderation also depends on how collaborative interactions are structured and on the social cues that participants encounter during the process. Our findings highlight that political identity signalling can meaningfully shape the benefits of collaboration, underscoring the importance of considering both political diversity and interaction design when developing community-based moderation systems.

\section*{Methods}

\subsection*{Data Collection}
In the summer of 2023, we recruited 432 active Prolific participants who self-identified as either Democrats or Republicans. We randomly paired them with partners from either the same or different political affiliations, resulting in three group configurations: (1) Democrat-Democrat (DD), (2) Republican-Republican (RR), and (3) Democrat-Republican (DR). Each team was randomly assigned to one of two treatments: in the ‘‘overt’’ treatment, participants could see each other’s political affiliations, while in the ‘‘covert’’ treatment, participants were unaware of each other’s affiliations. 

After pairing participants, we asked them to evaluate and write a contextualising note for one of 40 posts selected from either Democrat or Republican accounts (see Supplementary Information for more details). We instructed them to decide, based on the latest available evidence, whether the post was misinformed or potentially misleading. We asked them to explain their reasoning and include context to help others understand their judgment, and we encouraged them to cite external sources. These instructions closely followed those used in Community Notes at the time. After writing their initial note, we required them to chat with their partner and revise the note based on their shared consensus. Because collaboration was a core part of the design, we asked our participants to engage in discussion through the chat box. In the overt treatment, chat nicknames displayed political affiliation (e.g., “Democrat 1”, “Republican 2”), while in the covert treatment, nicknames were neutral (e.g., “Participant 1”,  “Participant 2”). Although participants in the covert condition could infer their partner’s political affiliation through conversation, this signalling remained indirect. At the end of the study, we asked participants to complete a brief survey about their reasoning process and overall experience (see Figure~\ref{FlowChart}). 

In total, our analysis includes 648 notes, split equally between collaborative and individual notes (216 each). Figure~S4 illustrates the breakdown of collaborative notes by group configuration and treatments. We discarded any interaction in which either participant ceased communication or in which chat text was submitted as a note. Any notes indicating that participants failed to reach an agreement or to comprehend the task were also discarded. 

To replicate the Community Notes helpfulness rating system, we recruited 1,610 additional participants from Prolific, evenly split between self-identified Democrats and Republicans (805 each). None of these participants had taken part in the first phase of the study. We asked them to evaluate the notes and rate their helpfulness on a scale from 0 to 10. In addition to the crowd-sourced assessments, the notes were also independently evaluated by three non-US expert PhD students. We instructed the experts to evaluate the notes objectively and avoid any political bias. To assess the reliability of expert evaluations, we computed the Intraclass Correlation Coefficient (ICC) using a two-way mixed-effects model, yielding 0.78, indicating substantial agreement among experts. 

\subsection*{Measurement}

On average, each note received ratings from five Democrats and five Republicans. We refer to the average rating from Democrats for a given note $a$ as the helpfulness score from Democrats ($H_{\rm D}^a$), and the average rating from Republicans as the helpfulness score from Republicans ($H_{\rm R}^a$). Experts also rated all notes, and we refer to the mean of their evaluations as the helpfulness score from experts ($H_{\rm E}^a$). We calculated these scores using the following equations:

\[
H_{\rm D}^a = \frac{1}{n_{\rm D}} \sum_{i=1}^{n_{\rm D}} h_{D_i}^a, \quad
H_{\rm R}^a = \frac{1}{n_{\rm R}} \sum_{i=1}^{n_{\rm R}} h_{R_i}^a, \quad
H_{\rm E}^a = \frac{1}{n_{\rm E}} \sum_{i=1}^{n_{\rm E}} h_{E_i}^a, 
\]

where $n_{\rm D}$, $n_{\rm R}$, and $n_{\rm E}$ denote the number of Democrats, Republicans, and experts who rated note $a$, respectively, and $h_{D_i}^a$, $h_{R_i}^a$, and $h_{E_i}^a$ refer to the evaluation by user $i$ on note $a$ from Democrats, Republicans, and experts, respectively.

To assess how collaboration affected note quality, we defined new variables $I_{\rm D}$, $I_{\rm R}$, and $I_{\rm E}$. They represent changes in helpfulness ratings from Democrats, Republicans, and experts, respectively. We calculated these indices as follows:
\[
I_{\rm D} = H_{\rm D}^{ab} - \frac{1}{2} \left( H_{\rm D}^a + H_{\rm D}^b \right),
\]
\[
I_{\rm R} = H_{\rm R}^{ab} - \frac{1}{2} \left( H_{\rm R}^a + H_{\rm R}^b \right),
\]
\[
I_{\rm E} = H_{\rm E}^{ab} - \frac{1}{2} \left( H_{\rm E}^a + H_{\rm E}^b \right),
\]
where $H_x^{ab}$ denotes the helpfulness score of the note co-authored by the team that originally wrote notes $a$ and $b$, and $x \in \{{\rm D},{\rm R}, {\rm E}\}$ refers to Democrats, Republicans, or experts. In cases where participants submitted two dissimilar notes, $H_x^{ab}$ was calculated as the average helpfulness score of the submitted notes. 

The description of other qualitative and quantitative measurements of the notes is detailed in Supplementary Information.

\subsection*{Statistical Analysis}

The $p$-values reported in the main text are based on parametric independent two-sample $t$-tests; corresponding non-parametric bootstrap tests are reported in the Supplementary Information. Parametric comparisons were conducted using independent two-sample $t$-tests. As a non-parametric robustness check, we used a bootstrap resampling procedure. For each comparison, observations were sampled with replacement within each group while preserving the original group sizes. We then calculated the difference in resampled group means. For directional hypotheses, the bootstrap p-value was computed as the proportion of bootstrap iterations in which the resampled mean difference was in the direction opposite to the hypothesised effect. For example, when testing whether collaboratively written notes were more helpful than individually written notes, we calculated the proportion of bootstrap samples in which the mean helpfulness score for individual notes exceeded that for collaborative notes. We used 10,000 bootstrap iterations for the main comparisons and 100,000 iterations for subgroup analyses. All tests were conducted in Python using \texttt{scipy.stats}.

To assess whether the distributions of note helpfulness contained distinct helpful and unhelpful modes, we calculated the sample-corrected Bimodality Coefficient (BC) separately for individually and collaboratively written notes within each evaluator group. The BC combines the skewness ($m_3$), excess kurtosis ($m_4$), and sample size ($n$) of a distribution:

\[
\mathrm{BC}
=
\frac{m_3^2 + 1}
{m_4 + 3\left(\frac{(n-1)^2}{(n-2)(n-3)}\right)},
\]
Following \citet{pfister2013good}, we compared each coefficient with the conventional benchmark of $5/9 \approx 0.555$. Values above this threshold are considered suggestive of bimodality, whereas lower values are more consistent with unimodality.

For regression analyses, we estimated ordinary least squares (OLS) models using the \texttt{lm()} function in R. Categorical predictors, including political-signalling condition, team partisan composition, and post type, were coded as factors with pre-specified reference categories. Specifically, the overt condition was used as the reference category for political signalling, RR teams were used as the reference category for team composition, and Republican posts were used as the reference category for post type. as the reference category for team composition, and Republican posts Model fit was summarised using adjusted $\rm R^2$ and F-statistics.

We also constructed two additional measures: political diversity and post alignment. Political diversity is a binary variable indicating whether a team consisted of one Democrat and one Republican participant. DR teams were coded as politically diverse, while DD and RR teams were coded as aligned. Post alignment measures the number of participants in the team whose political affiliations aligned with the post's partisanship. Thus, DD teams evaluating Democrat posts and RR teams evaluating Republican posts were assigned a value of 2. Politically diverse teams (DR) were assigned a value of 1. DD teams evaluating Republican posts and RR teams evaluating Democrat posts were assigned a value of 0. We treated post-alignment as an ordinal variable.

\subsection*{Robustness and Manipulation Checks}

We conducted additional robustness and manipulation checks, which are reported in the Supplementary Information. First, to assess the robustness of the main results, we implemented a bootstrap procedure in which we repeatedly sampled subsets of posts and recalculated the mean helpfulness scores across the relevant team-composition and treatment categories. Specifically, we selected half of the Democrat and Republican posts in each iteration, grouped observations by team dynamics and treatment condition, and recalculated the category-level means. This procedure was repeated across all possible combinations of selecting 10 out of 20 posts, yielding 184,756 iterations. The resulting bootstrap distributions are reported in the Supplementary Information.

Second, to examine whether the overt political-signalling condition altered the interactional environment of collaboration, we conducted an LLM-assisted manipulation check using the chat messages. We used an LLM pipeline to calculate the partisan content salience, partisan identity salience, disagreement, and concessionary behaviour from the chat messages. This analysis was used to assess whether making political affiliation visible was associated with greater partisan salience or changes in collaborative dynamics during note writing. The full coding procedure and results are reported in the Supplementary Information.

\section*{Data Availability}
Data used in the study are available at \url{https://doi.org/10.5281/zenodo.17468607}.

\section*{Acknowledgements}
This publication has emanated from research supported in part by a grant from Taighde Éireann – Research Ireland under Grant numbers 18/CRT/6049 and IRCLA/2022/3217. TY acknowledges support from Workday, Inc. For the purpose of Open Access, the author has applied a CC BY public copyright licence to any Author Accepted Manuscript version arising from this submission.

\section*{Author Contribution}
GJ, SM, MS, and TY designed the experiments. GJ and SM conducted the experiments and analysed the data. SM, GJ, and TY drafted the paper. MS and TY secured the funding. TY supervised the project. All authors contributed to the writing of the manuscript and gave final approval for publication.

\section*{Competing Interest}
The authors declare no competing interests.

\section*{Correspondence}
Correspondence and requests for materials should be addressed to Taha Yasseri: taha.yasseri@tcd.ie.

\bibliographystyle{apacite}  
\bibliography{references}  

\clearpage
\renewcommand{\thetable}{S\arabic{table}}
\renewcommand{\thefigure}{S\arabic{figure}}

\setcounter{table}{0}  
\setcounter{figure}{0} 
\appendix
\section*{Supplementary Information for \\Overt Political Signalling Undermines Collective Intelligence in Community-Based Content Moderation\\ Gabriela Juncosa, Saeedeh Mohammadi, Margaret Samahita, and Taha Yasseri\\}

\subsection*{Note Writing Experiment} We designed a note-writing experiment to examine the effect of collaboration on contextual notes for political Social Media posts. We recruited 432 active U.S.-based Prolific participants who self-identified as Democrats or Republicans. At the start of the session, we verified their political affiliation by asking them to indicate how strongly they identified with the two major political parties in the United States. The exact wording of the question was: “How strongly do you identify with either of the two main political parties in the U.S. (Democratic and Republican parties)?” Participants responded using a 10-point scale, where 1 indicated “Strongly Democrat” and 10 indicated “Strongly Republican”. This measure was subsequently used to classify participants’ political affiliation for the matching procedure.

Participants were randomly paired with partners who either shared or differed in political affiliation, resulting in three group configurations: (1) Democrat–Democrat (DD), (2) Republican–Republican (RR), and (3) Democrat–Republican (DR). Each team was then randomly assigned to one of two experimental treatments. In the overt treatment, participants were informed of their partner’s political affiliation, whereas in the covert treatment, participants were not. This matching procedure yielded six distinct team types.

\begin{figure}[H]
  \centering
      \includegraphics[width=0.75\textwidth]{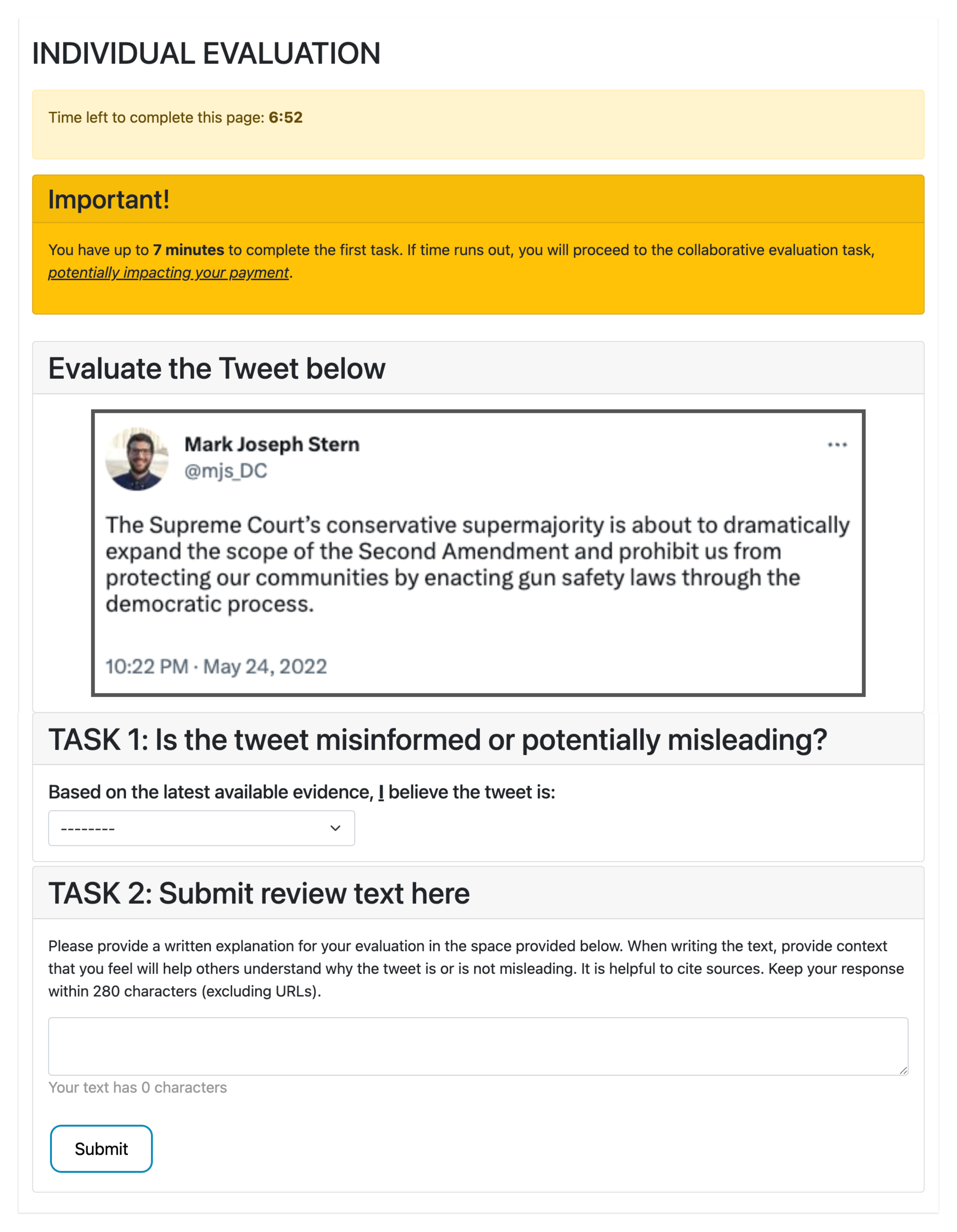}
\caption{Individual evaluation task interface. Participants reviewed a social media post and judged its credibility. They then provided a written explanation of up to 280 characters, with optional contextual details and source citations (URLs excluded from the character limit), following instructions modelled on Community Notes. Participants had up to seven minutes to complete the task before an attention check was triggered; those who passed were redirected to the collaborative evaluation phase.}
\label{fig:IndividualEval}
\end{figure}

Once teams were formed, each was randomly assigned to evaluate one of 40 available social media posts, comprising 20 pro-Democrat and 20 pro-Republican posts (see \textit{Protocol for selecting the social media posts} below). Our goal was to have each post evaluated at least once by all six treatment groups; in other words, we aimed to collect 240 valid evaluations. Participants were informed that the experiment consisted of two tasks--an individual and a collaborative evaluation. In the individual evaluation phase, participants were asked to review a social media post and assess its credibility. Specifically, they were prompted with the question: “Based on the latest available evidence, do you believe the tweet is misinformed or potentially misleading?” In addition, participants were instructed to explain their reasoning in writing by submitting a text limited to 280 characters. They were encouraged to provide any context that could help others understand why the post was or was not misleading and to cite relevant sources; however, URLs were not included in the character count. These instructions closely followed those used in Community Notes. Participants had up to seven minutes to complete the individual evaluation. Failure to do so triggered an attention check; if the check was passed, participants were then redirected to the collaborative evaluation task. The final dataset ($N=216$) includes only submissions that contain both individual and collaborative notes. Figure~\ref{fig:IndividualEval} provides a screenshot illustrating the interface used for the individual evaluation.

Before proceeding to the collaborative evaluation, participants who judged the social media post to be misleading were asked to provide additional insight into their reasoning by responding to Questions 1–5 below. Participants who judged the post to be \textit{not} misleading were asked to respond to Question 5 only. 
\begin{enumerate}
    \item Why do you believe the tweet may be misleading? Choose all that apply.
    \begin{itemize}
        \item It contains a factual error
        \item It contains manipulated media
        \item It contains outdated information that may be misleading 
        \item It is a misrepresentation or missing important context 
        \item It presents a disputed claim as a fact 
        \item It is satire/a joke that may be misinterpreted
        \item Other
    \end{itemize}
    \item If the tweet gained widespread exposure, who would likely believe its message?
    \begin{itemize}
        \item Few
        \item Many
    \end{itemize}
    \item If many believed the tweet, it might cause:
    \begin{itemize}
        \item Little harm 
        \item Considerable harm 
    \end{itemize}
    \item Finding and understanding the correct information would be:
    \begin{itemize}
        \item Easy
        \item Challenging
    \end{itemize}
    \item Have you done any of the following? Choose all that apply.
    \begin{itemize}
        \item Editing Wikipedia.
        \item Leaving a review for a business on Yelp, Google Maps, etc. 
        \item Fact-checking online content. 
        \item Leaving a comment on a news website.
        \item Contributed to any other user-generated content. 
    \end{itemize}
\end{enumerate}

\begin{figure}[H]
  \centering
      \includegraphics[width=0.65\textwidth]{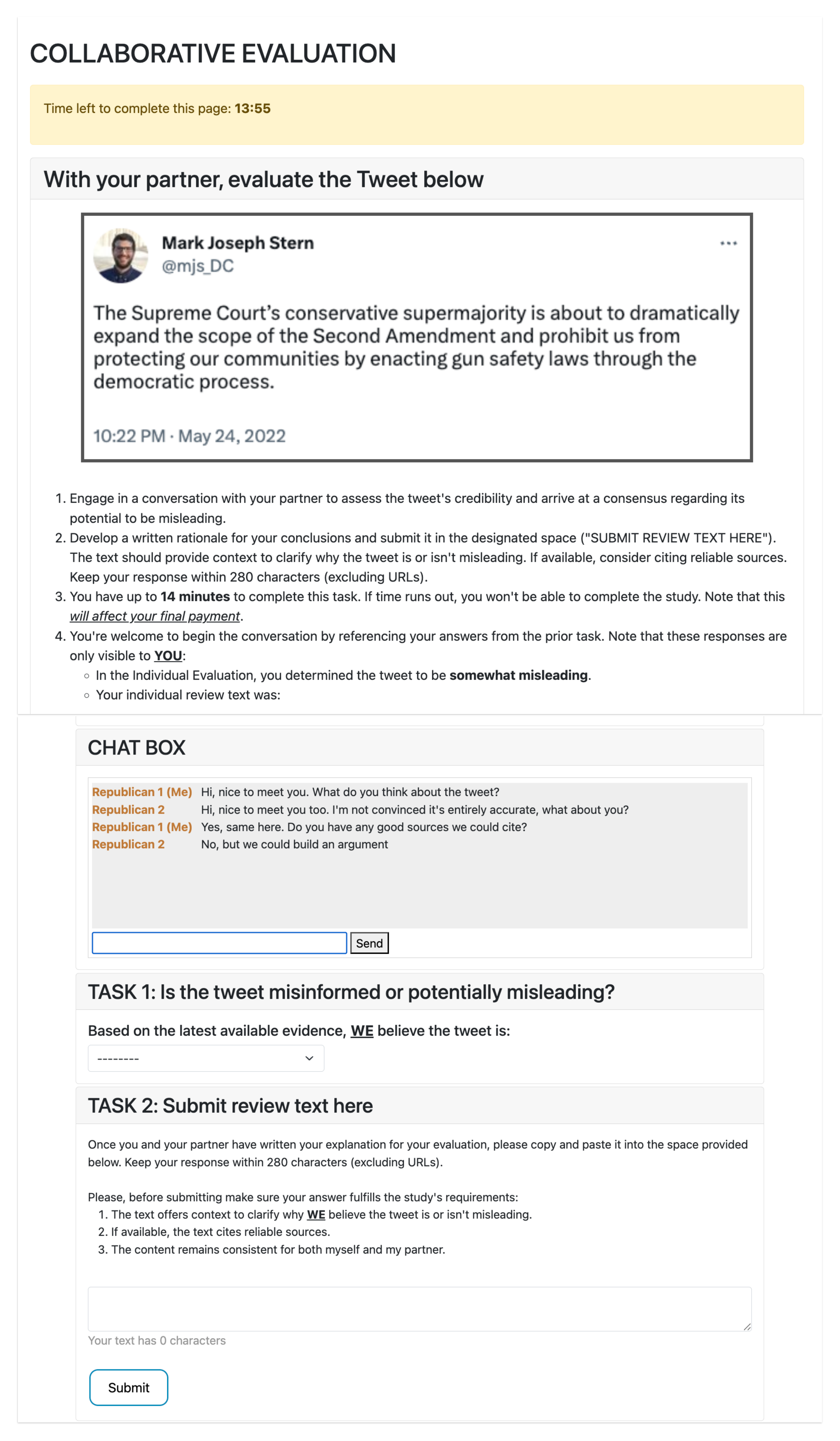}
\caption{Collaborative evaluation task interface. Teams jointly reviewed a social media post and assessed its credibility by writing a shared note that met the same requirements as the individual evaluation. Participants were instructed to actively collaborate while composing their explanations, with an integrated chat box monitored for compliance and task quality. In the overt condition, chat nicknames displayed political affiliation (e.g., “Democrat 1,” “Republican 2”), whereas in the covert condition, nicknames were neutral (e.g., “Participant 1,” “Participant 2”). Teams had up to thirteen minutes to complete the collaborative evaluation.}
\label{fig:CollaborativeEval}
\end{figure}

After completion of the individual evaluation, participants were asked to repeat the exercise, this time working with their assigned partner. Specifically, they were instructed to submit the collaborative evaluation with the same requirements as before—to determine whether the social media post was misinformed or potentially misleading, and to compose a joint explanation. Participants were required to collaborate actively while writing the note, as collaboration was a core element of the study design. To facilitate this process, we provided a chat box and monitored it for compliance and to assess task quality. In the overt treatment, chat nicknames displayed participants’ political affiliation (e.g., “Democrat 1,” “Republican 2”), whereas in the covert treatment, nicknames were neutral (e.g., “Participant 1,” “Participant 2”). Teams had up to thirteen minutes to complete the collaborative evaluation. Figure~\ref{fig:CollaborativeEval} provides a screenshot illustrating the interface used for the collaborative evaluation.

After the collaborative evaluation was submitted, we checked whether the two participants’ texts matched. If they did, participants proceeded to complete a brief survey about their reasoning process and overall experience. Specifically, they were asked: \textit{“How easy was it for you and your partner to come up with a text you both agreed on?”} The response options were: \textit{Extremely easy, Somewhat easy, Somewhat difficult, and Extremely difficult}. 

If the texts did not match, participants were given a second opportunity to review their submission. If they still could not reach an agreement, they were asked to provide additional information about their experience. In particular, they were presented with one or both of the following questions, depending on the specific mismatch criteria:
\begin{enumerate}
    \item Why do you believe your evaluation labels (i.e., misleading/not misleading) do not match?
    \begin{itemize}
        \item  We had different opinions and couldn't agree on a label
        \item My partner was unresponsive
        \item I believe this is a mistake but I couldn't go back to the task to review the label
        \item Other
    \end{itemize}
    \item Why do you believe your evaluation texts do not match?
    \begin{itemize}
        \item We had different opinions and couldn't agree on a text
        \item My partner was unresponsive
        \item I believe this is a mistake but I couldn't go back to the task to review the text
        \item I don't believe the texts to be significantly different. Both convey the same message
        \item Other
    \end{itemize}
\end{enumerate}

\begin{figure}[H]
  \centering
      \includegraphics[width=\textwidth]{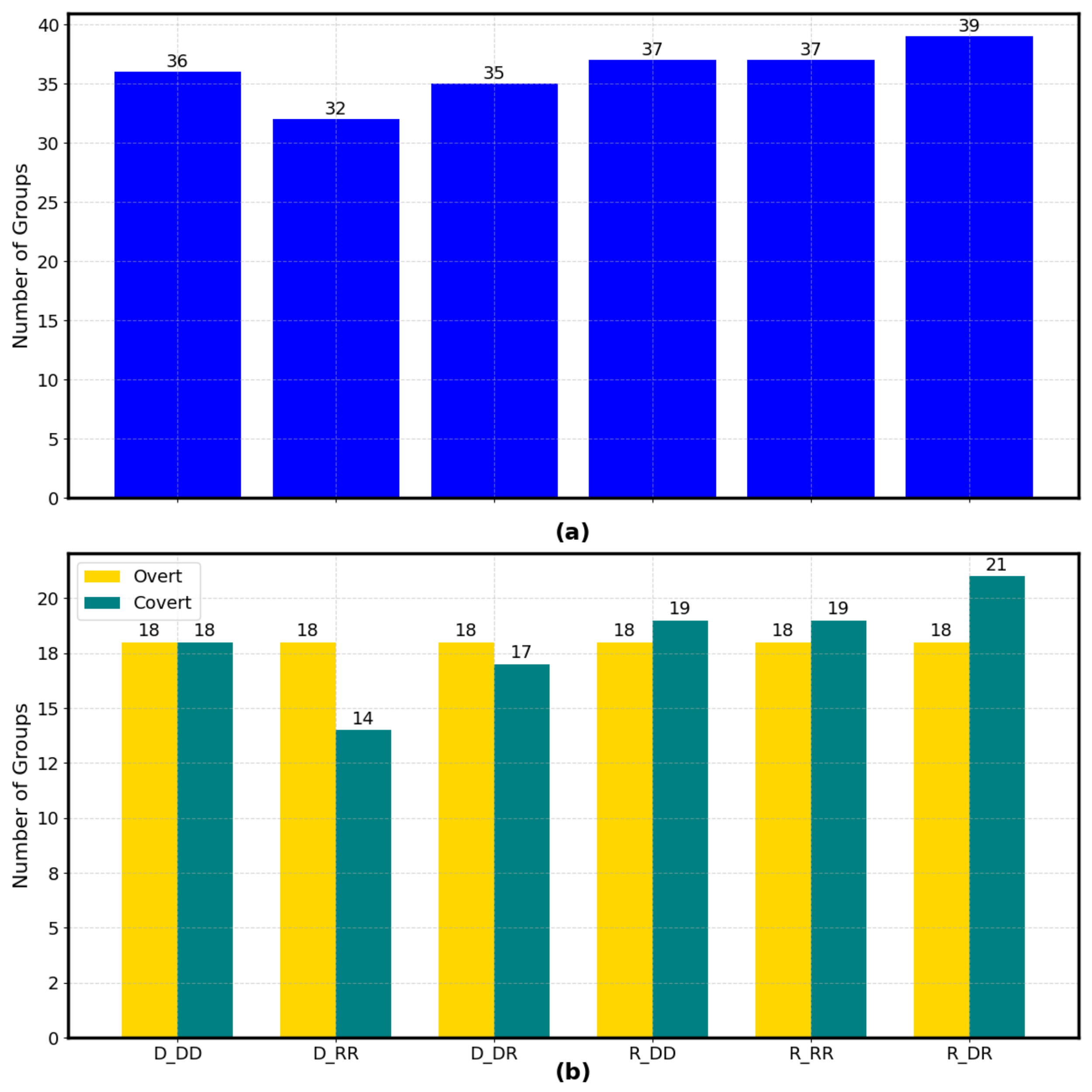}
\caption{Breakdown of the total number of collaborative notes ($N=216$) by (a) group configuration and (b) political identity signalling condition. The notation of the x-axis follows this format: "post partisanship\_team partisanship." For example, if a Democrat and a Republican evaluated a Democrat post, they would be labelled as D\_DR. Similarly, if two Republicans evaluated a Republican post, they would be labelled as R\_RR.}
\label{fig:CountsBreakdowns}
\end{figure}

We wanted each post evaluated at least once by each of the six treatment groups, aiming for a minimum of 20 completed evaluations per category. After the experiment, we carefully reviewed all data to ensure that minimum quality standards were met. The final dataset includes 216 valid evaluations. Figure \ref{fig:CountsBreakdowns} panel (a) presents the combined totals for the overt and covert treatments, while panel (b) reports the disaggregated numbers of accepted notes by treatment category. The x-axis notation follows the format “post partisanship\_team partisanship.” For example, the number of accepted notes on social media posts from Republican sources, evaluated by a team of two Democrats, is denoted by R\_DD. In several cases, the number of evaluations fell slightly below the 20-evaluation target due to variations in participant availability during recruitment and quality-control removals. At the time of data collection, Prolific had a larger pool of active participants who identified as Democrats (approximately 10,000) compared with Republicans (approximately 3,000). To minimise wait times, we occasionally paired participants with others who were active and available.

\subsection*{Note Evaluation Experiment}
We designed the note evaluation experiment to assess the helpfulness and quality of the notes written in the main study. We recruited participants on Prolific and asked them to evaluate the notes. After entering the study, participants read an information sheet outlining the study’s purpose, ethical approval, and potential risks and benefits. They then provided consent and entered their Prolific ID.

Next, we asked participants to indicate their political partisanship on a nine-point scale, matching the procedure used in the note-writing experiment.

In the main task, we showed each participant a social media post accompanied by 10 notes about it. We instructed them to evaluate each note based on the following criteria:
\begin{itemize}
    \item \textbf{Context}: Does this note provide sufficient context for the post, equipping the reader with enough information?
    \item \textbf{Fairness}: Is this note fair, impartial, and unbiased in tone?
    \item \textbf{Relevance}: Does the note directly address the issues raised in the post, or is it mere speculation or personal opinion unrelated to the post’s claims?
    \item \textbf{Readability}: Is this note well-written, free from typos and grammatical errors, and can be easily understood?
    \item \textbf{Sources}: Are there any sources cited in the note, and are they reliable and supportive of the post's argument?
\end{itemize}

\begin{figure}[H]
  \centering
      \includegraphics[width=0.8\textwidth]{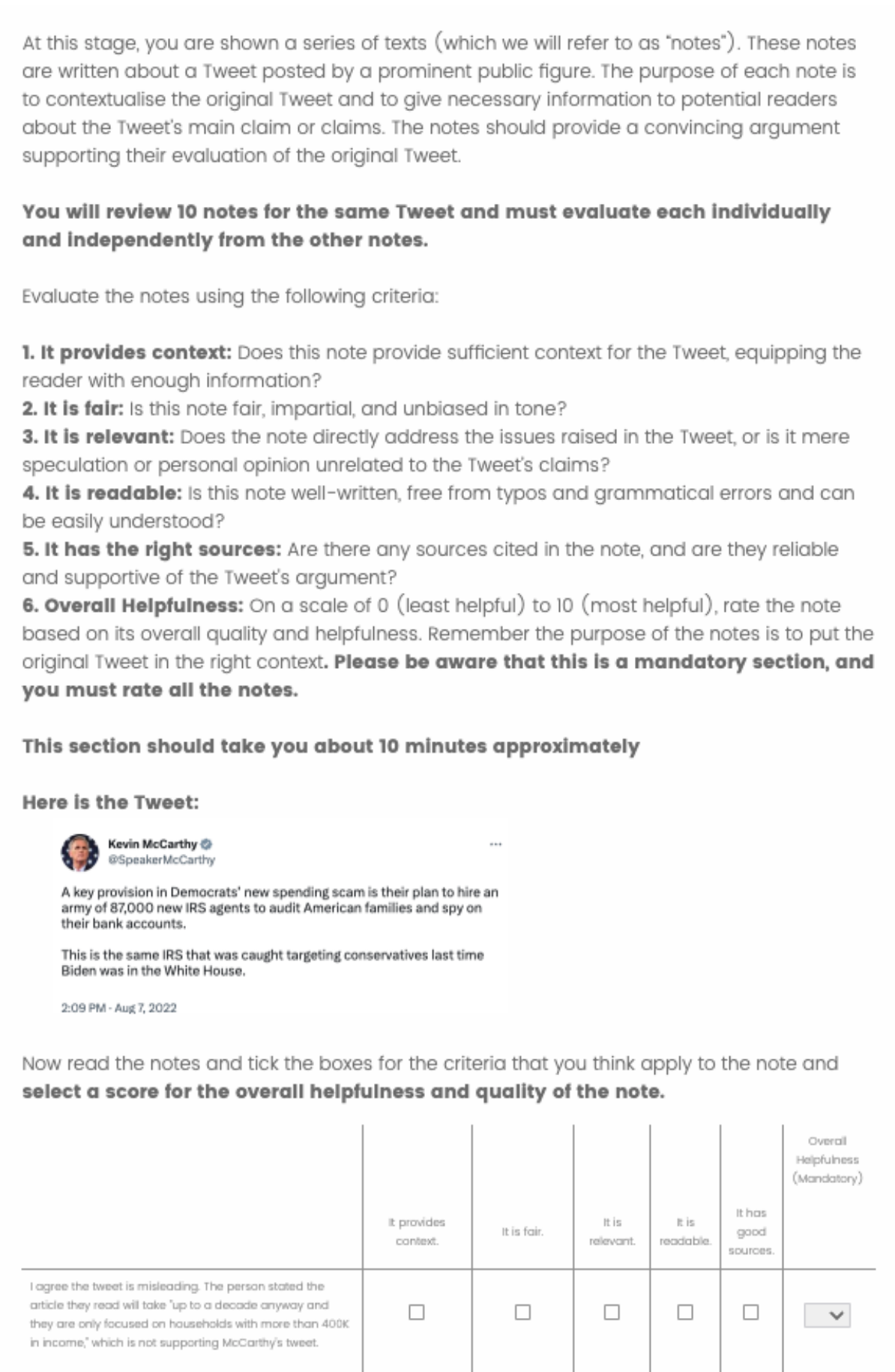}
\caption{A screenshot of the description provided to participants for evaluating the notes.
}
  \label{fig:note_evals}
\end{figure}

Finally, participants rated each note’s overall helpfulness on a scale from 0 to 10. Figure~\ref{fig:note_evals} presents an example of the instructions and a sample post shown during the evaluation task.

\begin{figure}[H]
  \centering
      \includegraphics[width=\textwidth]{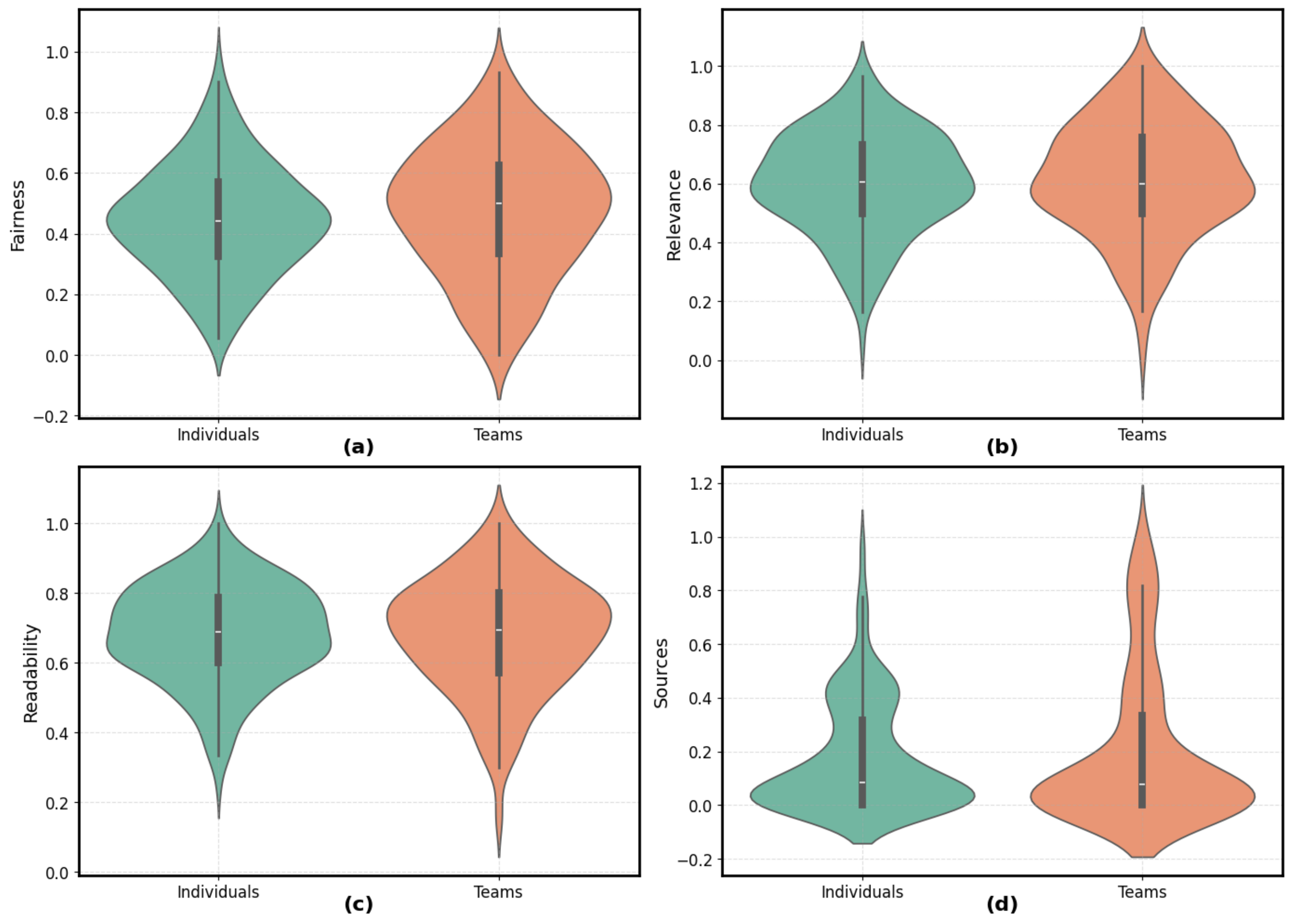}
\caption{The distribution of the qualitative measures for notes written by individuals and teams based on ratings from evaluators who self-identified as Democrats. (a) The distribution of the fairness scores given to the note. (b) The distribution of the relevance scores given to the note. (c) The distribution of the readability scores given to the note. (d) The distribution of the reliability score of the sources' given to the note.}
\label{fig:inds_groups_qualitative_d}
\end{figure}

\begin{figure}[H]
  \centering
      \includegraphics[width=\textwidth]{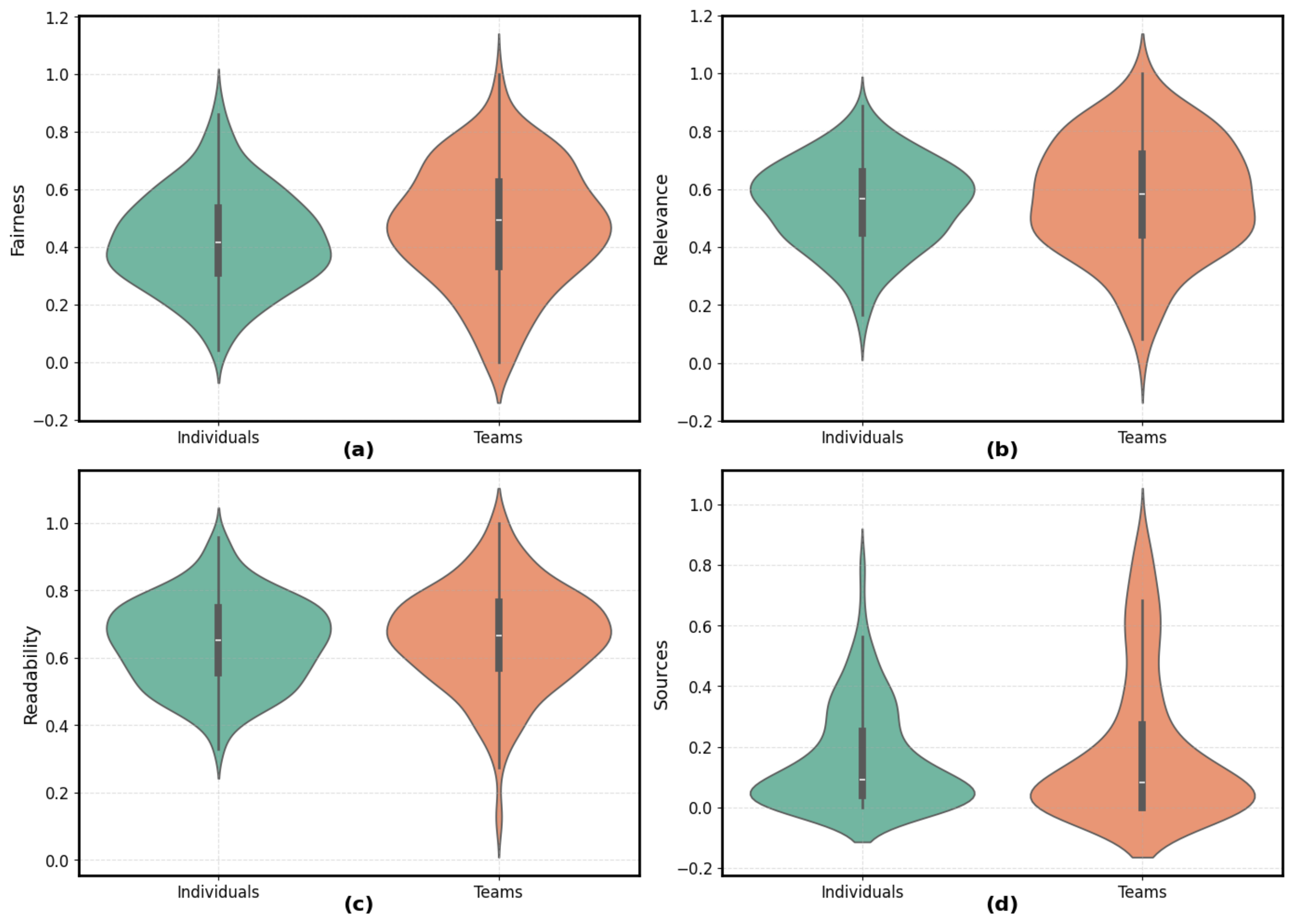}
\caption{Distribution of qualitative measures for notes written by individuals and teams, based on ratings from evaluators who self-identified as Republicans. (a) The distribution of the fairness scores given to the note. (b) The distribution of the relevance scores given to the note. (c) The distribution of the readability scores given to the note. (d) The distribution of the reliability of the sources' scores given to the note.}
\label{fig:inds_groups_qualitative_r}
\end{figure}

\begin{figure}[H]
  \centering
      \includegraphics[width=\textwidth]{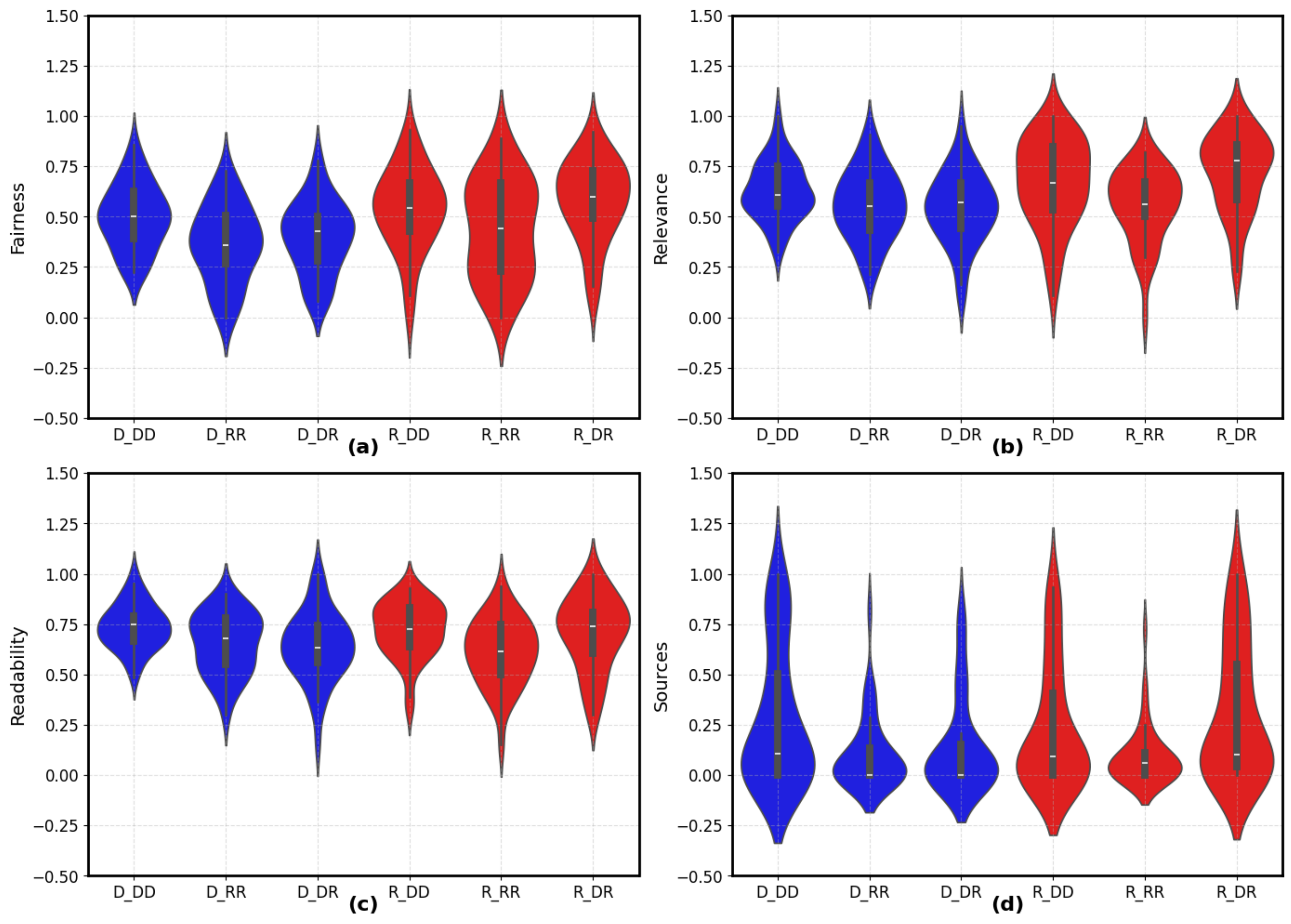}
\caption{The distribution of the qualitative measures for notes across different group configurations based on ratings from evaluators who self-identified as  Democrats. (a) The distribution of the fairness scores given to the note. (b) The distribution of the relevance scores given to the note. (c) The distribution of the readability scores given to the note. (d) The distribution of reliability scores for the sources given to the note. The notation on the x-axis follows the format "post partisanship\_team partisanship." For example, if a Democrat and a Republican evaluated a Democrat post, they would be labelled as D\_DR. Similarly, if two Republicans evaluated a Republican post, they would be labelled as R\_RR.}
\label{fig:groups_qualitative_d}
\end{figure}

\begin{figure}[H]
  \centering
      \includegraphics[width=\textwidth]{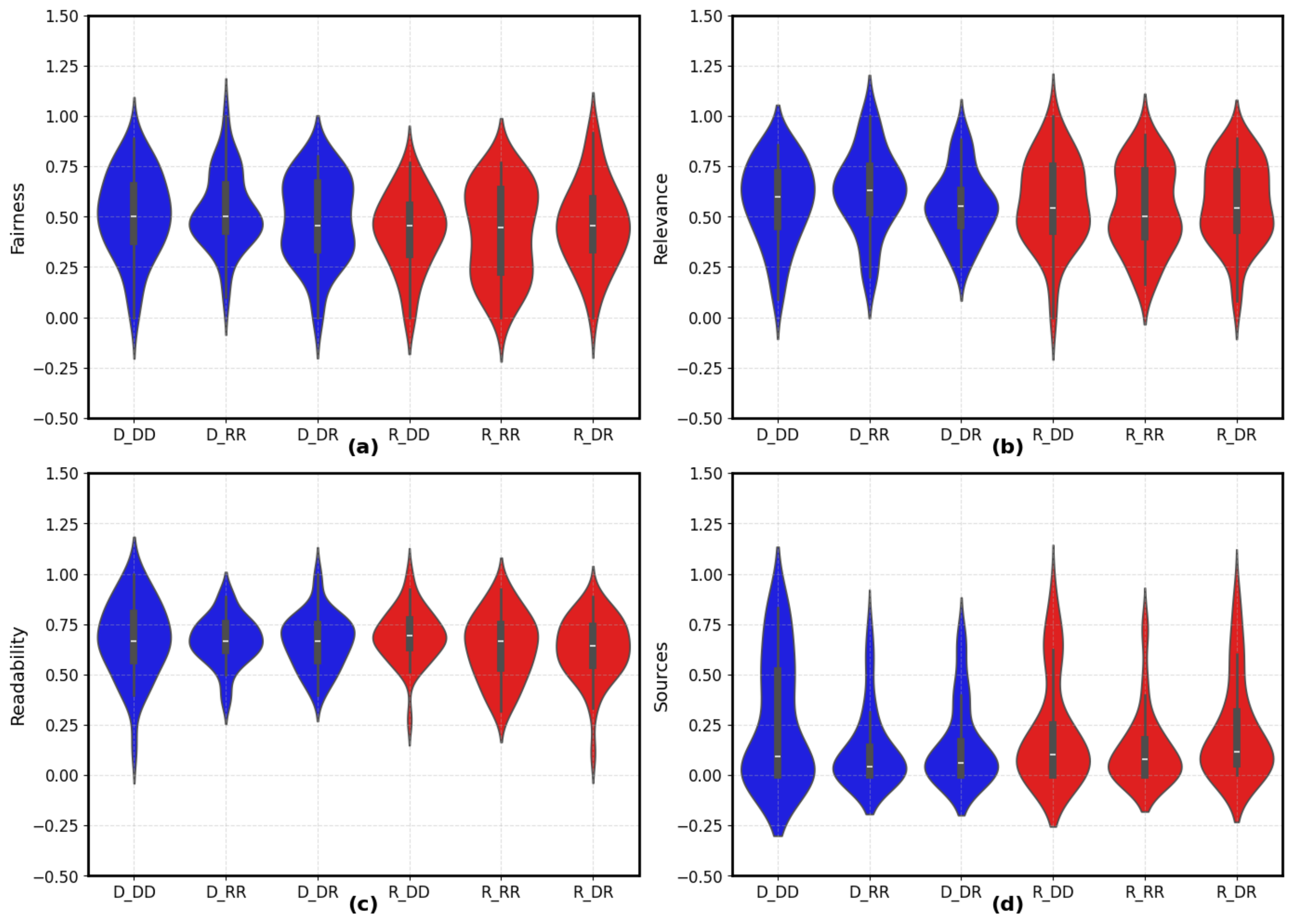}
\caption{Distribution of qualitative measures for notes across group configurations, based on ratings from evaluators who self-identified as Republicans. (a) The distribution of the fairness scores given to the note. (b) The distribution of the relevance scores given to the note. (c) The distribution of the readability scores given to the note. (d) The distribution of reliability scores for the sources given to the note. The notation on the x-axis follows the format "post partisanship\_team partisanship." For example, if a Democrat and a Republican evaluated a Democrat post, they would be labelled as D\_DR. Similarly, if two Republicans evaluated a Republican post, they would be labelled as R\_RR.}
\label{fig:groups_qualitative_r}
\end{figure}

Figures~\ref{fig:inds_groups_qualitative_d}, \ref{fig:inds_groups_qualitative_r}, \ref{fig:groups_qualitative_d}, and \ref{fig:groups_qualitative_r} show the distribution of these metrics.

\subsection*{Protocol for selecting the Social Media Posts}
We selected 40 posts from \url{X.com} for the note-writing experiment, using data drawn from the Community Notes dataset. We followed a set of predefined criteria to ensure that the posts were relevant and comparable across political contexts:

\begin{itemize}
\item Each post contained at least one of the following keywords: Democrat or Republican.
\item Each post had at least two notes tagged as “misinformed or potentially misleading” in the Community Notes dataset.
\item The content of the post was related to the United States.
\item The text was clear and easy to understand.
\item The post did not include a video or an external link required for comprehension.
\end{itemize}

After compiling the eligible posts, we manually classified them as either pro-Democrat or pro-Republican based on their content. The final set included 20 pro-Democrat posts and 20 pro-Republican posts.

\subsection*{Quantitative Analysis of the Notes}

To gain a more comprehensive understanding of the notes, we extracted several quantitative metrics. The metrics are defined as follows:

\begin{itemize}
  \item \textbf{Length}: The length of the note, measured by the number of words.
  \item \textbf{Links}: The number of hyperlinks included in the note as references.
  \item \textbf{Sophistication Level}: A standard score indicating readability, complexity, and grade level (according to the U.S. education system). This score is calculated using the `textstat` library (version 0.7.3) in Python (version 3.11).
  \item \textbf{Stats}: A binary variable indicating the presence or absence of numerical data in the text.
\end{itemize}

\begin{figure}[H]
  \centering
      \includegraphics[width=\textwidth]{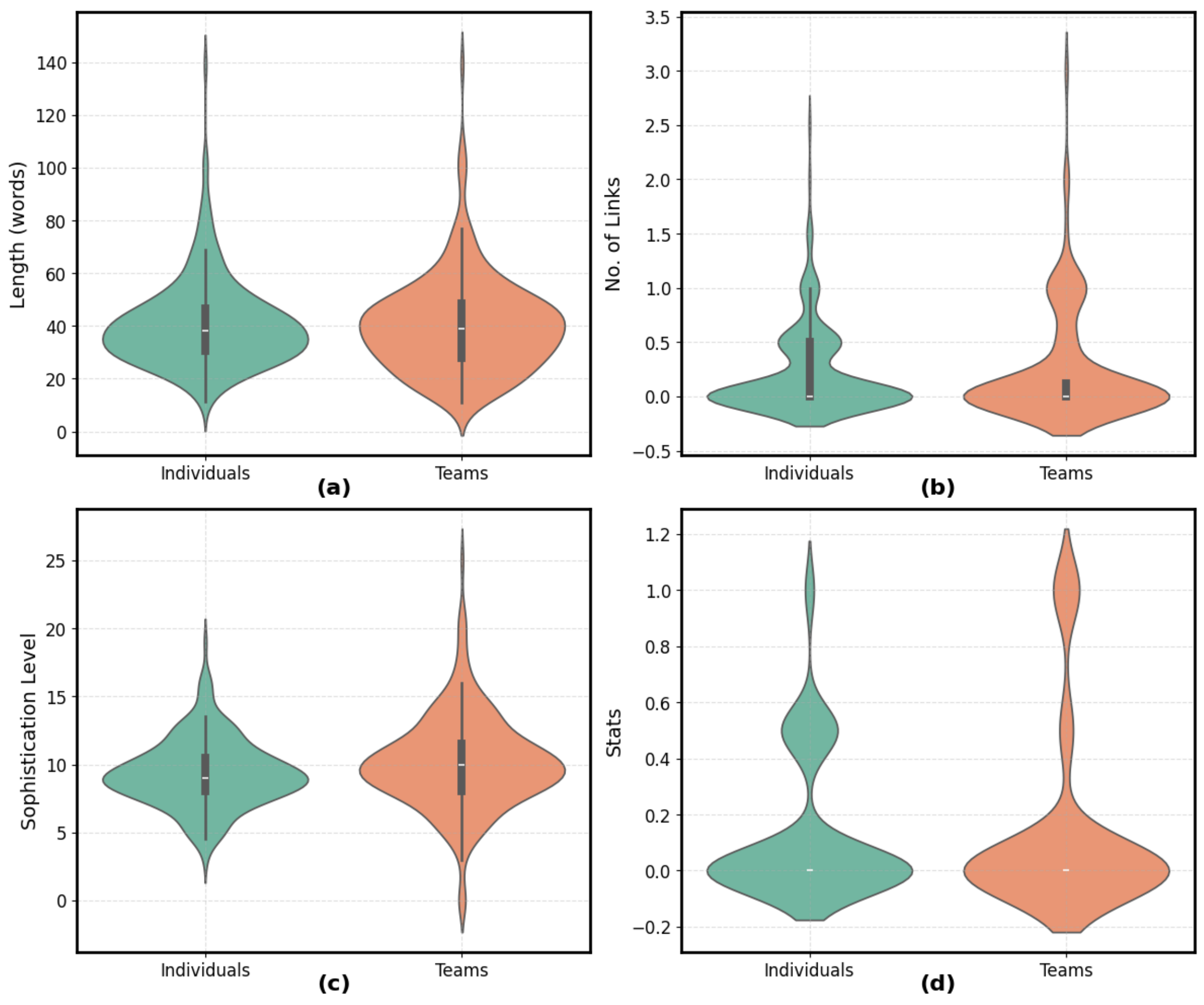}
\caption{The distribution of the quantitative measures for notes written by individuals and teams. (a) Length of the notes by words. (b) Number of links used in the notes. (c) The sophistication level of the notes. (d) The likelihood of the existence of numbers in a note.}
\label{fig:inds_groups_quants}
\end{figure}

\begin{figure}[H]
  \centering
      \includegraphics[width=\textwidth]{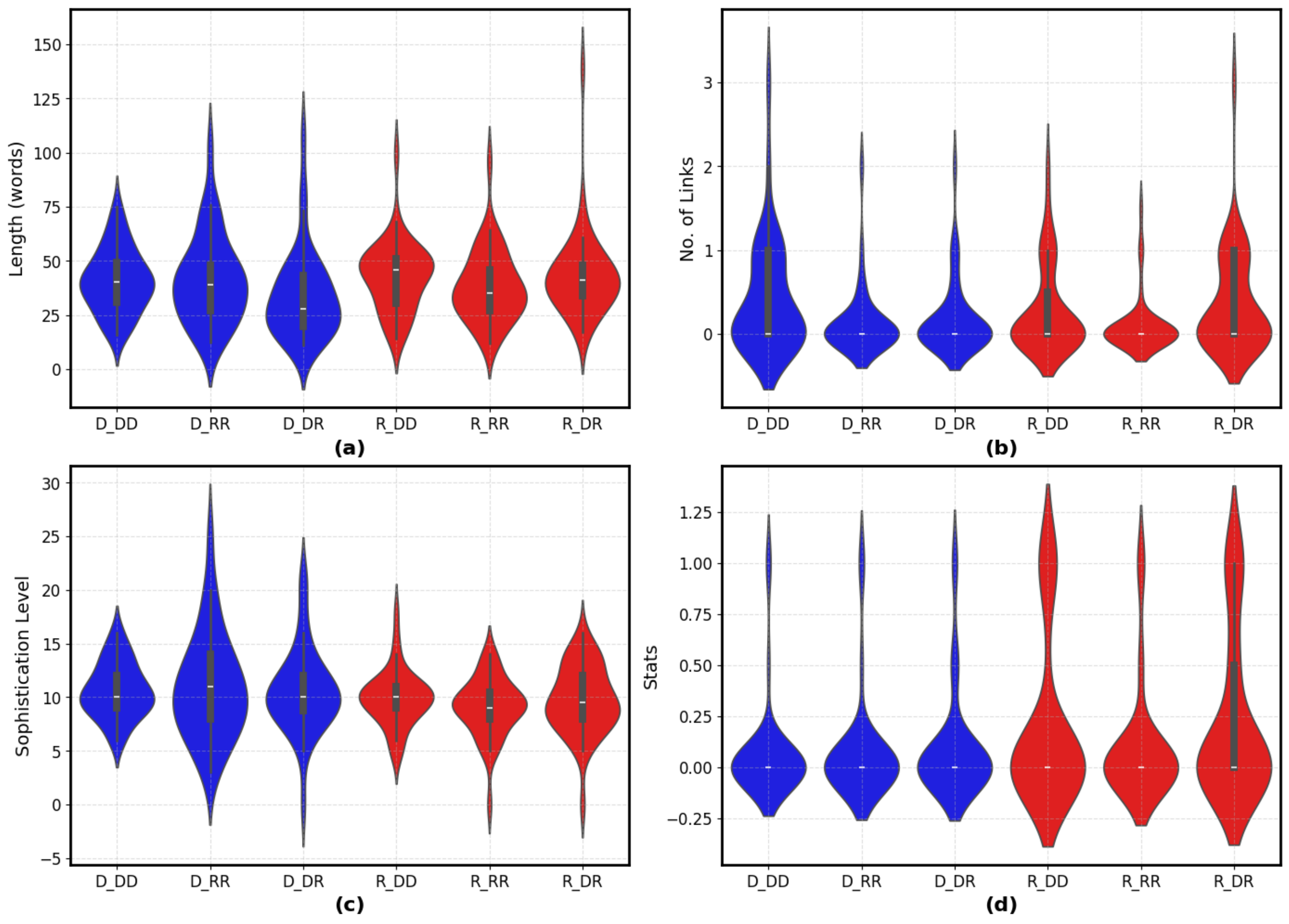}
\caption{The distribution of the quantitative measures for notes across different group configurations. (a) Length of the notes by words. (b) Number of links used in the notes. (c) The sophistication level of the notes. (d) The likelihood of the existence of numbers in a note.}
\label{fig:groups_quants}
\end{figure}

Figures~\ref{fig:inds_groups_quants} and \ref{fig:groups_quants} show the distributions of the metrics for each category.

The change in each of these metrics was calculated using the following equation:

\[
I_{\rm X} = X^{ab} - \frac{1}{2} \left( X^a + X^b \right),
\]

where \( X \) represents any of the aforementioned metrics. \( X^{ab} \) denotes the metric for the note co-authored by teams \( a \) and \( b \), while \( X^a \) and \( X^b \) refer to the metrics for the notes written individually by teams \( a \) and \( b \), respectively.

We then performed a linear regression analysis using ordinary least squares (OLS), with $I$ as the independent variable and evaluations by Democrats, Republicans, and experts as the dependent variables. Table~\ref{tab:quant_regression} presents the results.

\subsection*{Bootstrap}
\begin{figure}[H]
  \centering
      \includegraphics[width=0.6\textwidth]{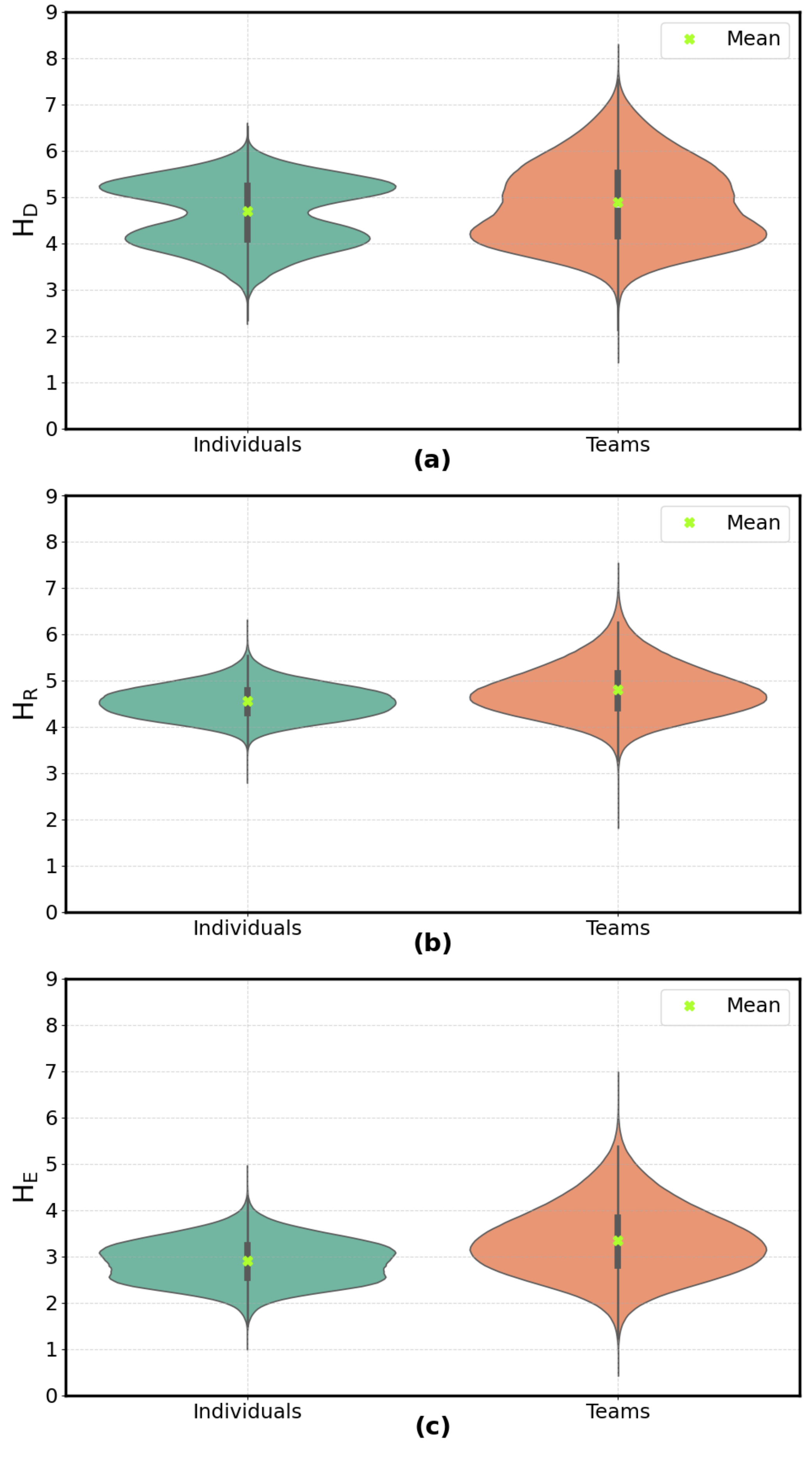}
\caption{The distribution of helpfulness scores from the bootstrapping process in notes written by teams and individuals. The green x marks the mean of the original data points. (a) The helpfulness scores as evaluated by self-identified Democrats. (b) The helpfulness scores as evaluated by self-identified Republicans. (c) The helpfulness scores, as evaluated by experts. }
\label{fig:bootstrap_inds_groups}
\end{figure}

\begin{figure}[H]
  \centering
      \includegraphics[width=0.56\textwidth]{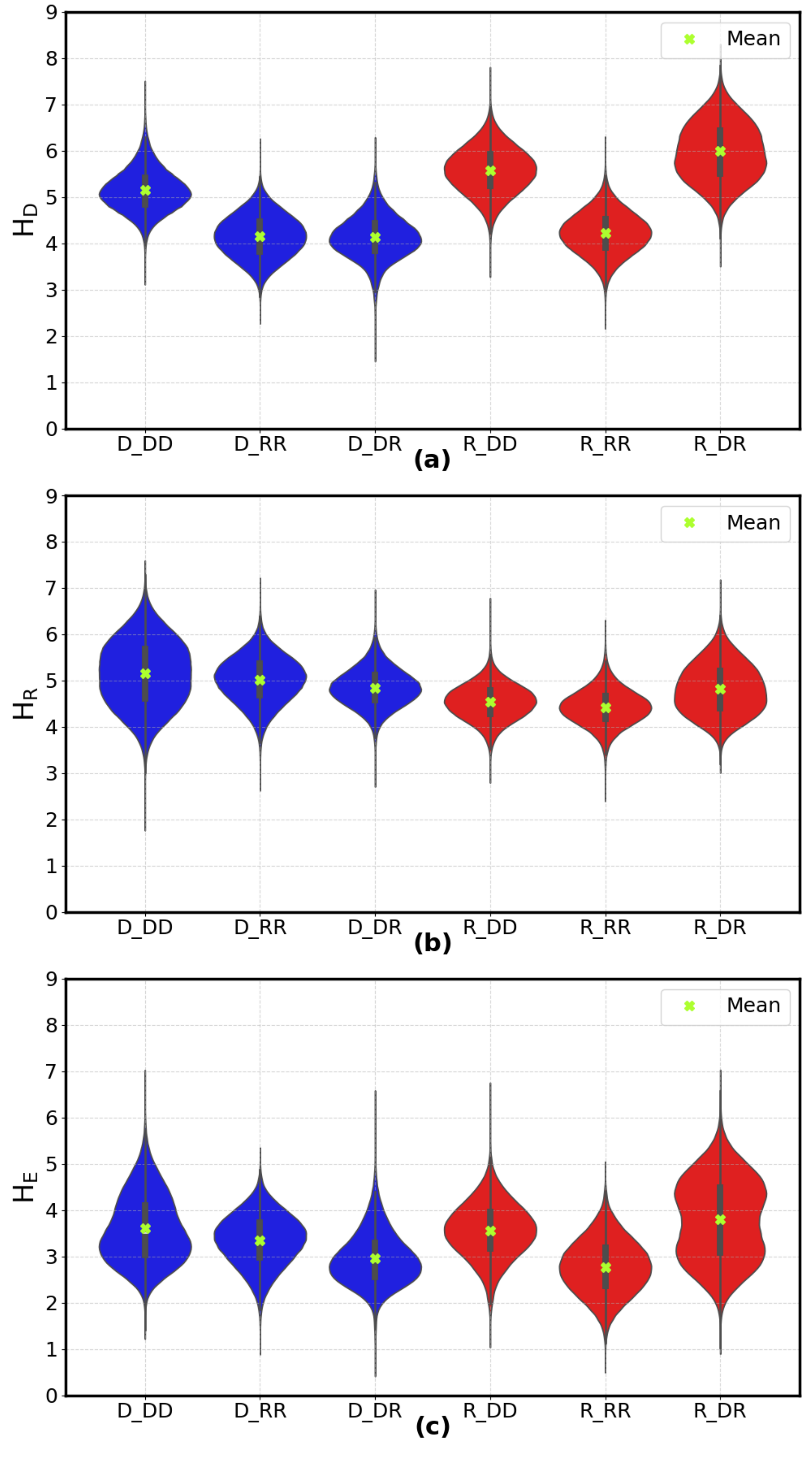}
\caption{The distribution of helpfulness scores from the bootstrapping process in notes written by teams across different group configurations. The green x marks the mean of the original data points. (a) The helpfulness scores as evaluated by self-identified Democrats. (b) The helpfulness scores as evaluated by self-identified Republicans. (c) The helpfulness scores, as evaluated by experts. The notation on the x-axis follows the format "post partisanship\_team partisanship." For example, if a Democrat and a Republican evaluated a Democrat post, they would be labelled as D\_DR. Similarly, if two Republicans evaluated a Republican post, they would be labelled as R\_RR.}
\label{fig:bootstrap_groups}
\end{figure}

\begin{figure}[H]
  \centering
      \includegraphics[width=0.6\textwidth]{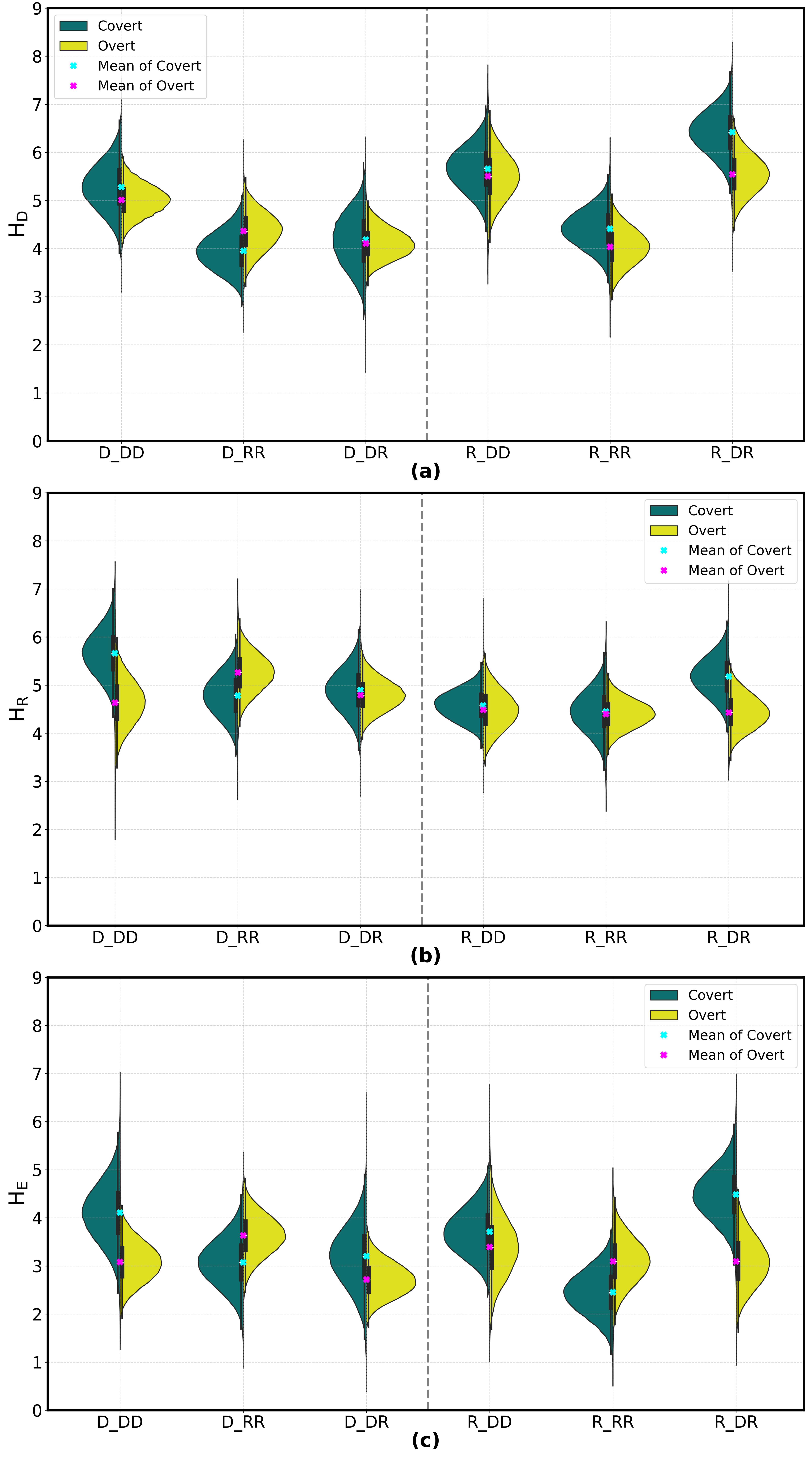}
\caption{The distribution of helpfulness scores from the bootstrapping process in notes written by teams, for covert and overt treatments across different group configurations. The teams in the covert and overt treatments are shown separately. The blue x marks the mean of the original data points in the covert treatment. The pink x marks the mean of the original data points in the overt treatment. (a) The helpfulness scores as evaluated by self-identified Democrats. (b) The helpfulness scores as evaluated by self-identified Republicans. (c) The helpfulness scores, as evaluated by experts. The notation on the x-axis follows the format "post partisanship\_team partisanship." For example, if a Democrat and a Republican evaluated a Democrat post, they would be labelled as D\_DR. Similarly, if two Republicans evaluated a Republican post, they would be labelled as R\_RR.}
\label{fig:bootstrap_covert_overt}
\end{figure}

To examine the robustness of the results, we employed the following bootstrapping method. First, we randomly selected half of the posts from both the Democrat and Republican datasets. Next, we categorised the data points by team dynamics and treatment. We then calculated the mean of the helpfulness scores from each resulting distribution. This process was repeated 184,756 times, corresponding to the number of possible combinations when selecting 10 options from 20. Figures \ref{fig:bootstrap_inds_groups}, \ref{fig:bootstrap_groups}, and \ref{fig:bootstrap_covert_overt} show the distribution of means for each category.

\subsection*{Manipulation Check on Chat Messages}
We conducted an LLM-assisted qualitative coding of the chat messages to assess whether the overt condition increased the salience of partisanship during interaction, and to examine interactional mechanisms that may explain differences in note quality. Each transcript consisted of the post under discussion, the messages sent by Participant 1, and the messages sent by Participant 2. The post under discussion, the messages sent by Participant 1, and the text of the post were provided to the model only as contextual information; all coding decisions were based on the participants’ chat messages.

We coded four transcript-level variables:

\begin{itemize}
    \item \textbf{Partisan content salience:} The extent to which participants discussed parties, ideological groups, partisan actors, or partisan interpretations.
    \item \textbf{Partisan identity salience:} The extent to which participants made their own or their partner’s political identity salient during the conversation.
    \item \textbf{Disagreement:} The extent to which participants challenged, contradicted, or argued against one another’s suggestions or interpretations.
    \item \textbf{Concessionary behaviour:} the extent to which participants acknowledged, incorporated, or accommodated the other participant’s input.
\end{itemize}

Together, these measures allow us to assess whether the overt treatment successfully increased the salience of partisan considerations during discussion relative to the covert treatment, providing a manipulation check for the experimental intervention, while also characterising how participants navigated differences of opinion when collaboratively drafting notes.

We used GPT-4.1 via the OpenAI API to score each transcript. To reduce conceptual overlap across coding tasks, we used two separate prompts. The first prompt measured partisan salience and produced two transcript-level scores: partisan content salience and partisan identity salience. The prompt used for the system message is provided below:

\begin{tcolorbox}[
  breakable,
  colback=gray!8,
  colframe=gray!50,
  title=\textbf{System message -- Partisan Salience},
  fonttitle=\bfseries,
  left=6pt, right=6pt, top=6pt, bottom=6pt,
  listing only,
  listing options={
    basicstyle=\ttfamily\footnotesize,
    breaklines=true,
    breakatwhitespace=true,
    columns=fullflexible
  }
]
You are an expert qualitative coder analysing chat transcripts from an academic study on collaborative content moderation.

You will be given:
1. The text of a tweet.
2. Messages sent by Participant 1.
3. Messages sent by Participant 2.

The participants are discussing whether the tweet is misleading and what note to produce.

Your task is to score the CHAT TRANSCRIPT, not the tweet itself, on two dimensions.

Dimension 1: Partisan content salience
Definition:
The extent to which participants discuss parties, ideological groups, partisan actors, or partisan interpretations as part of evaluating the tweet or writing the note.

Examples include references to Republicans, Democrats, liberals, conservatives, the left,
the right, political bias, partisan motives, or party-linked policy positions.

Important:
- Do NOT count partisan content that appears only in the tweet.
- Count only partisan content introduced or discussed by the participants themselves.
- This dimension does NOT require participants to reveal their own political identity.

Partisan content salience scale:
0 = No participant discussion of parties, ideological groups, or partisan framing.
1 = Minimal or passing participant reference to partisan content.
2 = Clear participant discussion of partisan content, but not central to the interaction.
3 = Frequent participant discussion of partisan content or partisan framing.
4 = Partisan content is dominant or central to the interaction.

Dimension 2: Partisan identity salience
Definition:
The extent to which participants make their own or the other participant's political identity, party affiliation, ideological side, or partisan group membership salient in the interaction.

Examples include:
- "I am Republican"
- "as a Democrat"
- "we Republicans"
- "your side"
- "our party"
- "people like us"
- "you liberals"
- asking the other participant about their party or ideology
- reacting to the other participant's revealed political identity

Important:
- Do NOT assign identity salience merely because participants discuss Republicans, Democrats, liberals, conservatives, or political bias.
- Partisan identity salience requires explicit or strongly implied reference to a participant's own or their partner's political identity.
- If participants discuss parties only as the subject of the tweet, identity salience
  should be 0.
- For this dimension, explicit self-identification is strong evidence even if it occurs only once.

Partisan identity salience scale:
0 = No participant political identity is mentioned or implied.
1 = Very indirect or ambiguous reference to participant political identity.
2 = Participant political identity is indirectly but clearly implied.
3 = One explicit reference to participant political identity, such as "we Republicans," "as a Democrat," "I'm conservative," or "your side."
4 = Participant political identity is mentioned repeatedly, used to frame disagreement, or becomes central to how participants negotiate the note.

General coding rules:
- Base scores only on participant messages, using the tweet only as context.
- Do not infer participants' political identities unless they explicitly state or strongly imply them.
- Each score should be supported by direct quotes from the chat.
- If there is no direct evidence for a dimension, assign 0.

\end{tcolorbox}

The second prompt measured interaction dynamics and produced two transcript-level scores: disagreement and concessionary behaviour. The full prompt used for the system message is shown below.

\begin{tcolorbox}[
  breakable,
  colback=gray!8,
  colframe=gray!50,
  title=\textbf{System message -- Interaction Dynamics},
  fonttitle=\bfseries,
  left=6pt, right=6pt, top=6pt, bottom=6pt,
  listing only,
  listing options={
    basicstyle=\ttfamily\footnotesize,
    breaklines=true,
    breakatwhitespace=true,
    columns=fullflexible
  }
]
You are an expert qualitative coder analysing chat transcripts from an academic study on collaborative content moderation.

You will be given:
1. The text of a tweet.
2. Messages sent by Participant 1.
3. Messages sent by Participant 2.

The participants are discussing whether the tweet is misleading and what note to produce.

Your task is to score the CHAT TRANSCRIPT, not the tweet itself, on two dimensions.

Dimension 1: Disagreement
Definition:
The extent to which Participant 1 and Participant 2 reject, challenge, contradict, criticise, or argue against each other's suggestions, interpretations, or proposed note content.

Important:
- Do NOT code disagreement with the tweet as disagreement.
- Do NOT code criticism of the tweet, political actors, or outside groups as disagreement.
- Disagreement means disagreement between Participant 1 and Participant 2.

Disagreement scale:
0 = No disagreement between participants.
1 = Minimal, mild, or indirect disagreement.
2 = Clear but limited disagreement.
3 = Frequent or clearly important disagreement.
4 = Strong, repeated, or central disagreement.

Dimension 2: Concessionary behaviour
Definition:
The extent to which one participant acknowledges, accepts, incorporates, revises, softens, or compromises in response to the other participant's suggestion, correction, challenge, or different viewpoint.

Important:
- Do NOT code simple agreement as concession.
- Phrases such as "good point," "you're right," "I see what you mean," "let's use your version," or revising a proposed note after the other participant's input are evidence of concession.

Concessionary behaviour scale:
0 = No concessionary behaviour.
1 = Minimal acknowledgement, adoption, or softening.
2 = Clear but limited concession, compromise, or incorporation of the other's input.
3 = Frequent or clearly important concessionary behaviour.
4 = Concession, compromise, or mutual accommodation is central to the interaction.
\end{tcolorbox}

For each transcript, the model returned structured JSON output containing the score for each dimension, a brief explanation, and direct quotations from the chat supporting the score. We instructed the model to base scores only on the participant messages, using the post only as context, and to assign a score of 0 when there was no direct evidence for a given dimension.

To improve reliability, we repeated the LLM coding procedure seven times for each transcript. The API call used temperature $= 0$ and requested JSON-formatted output. For each dimension, we calculated a vote-based score, defined as the uniquely most frequent score across the seven runs. If there was no unique most frequent score, the vote-based score was coded as missing. This occurred in only two cases across the dataset.

Finally, we assessed consistency across repeated LLM codings by calculating Fleiss' $\kappa$ across the seven runs for each coding dimension. Agreement was high for all measures: partisan content salience ($\kappa = 0.891$), partisan identity salience ($\kappa = 0.856$), disagreement ($\kappa = 0.914$), and concessionary behaviour ($\kappa = 0.821$), indicating substantial to near-perfect consistency across repeated applications of the coding rubric.

\begin{figure}[H]
  \centering
      \includegraphics[width=\textwidth]{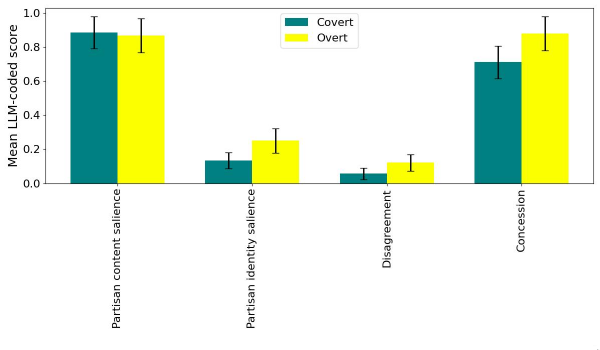}
\caption{LLM-coded Chat Characteristics by Covert and Overt condition. Bars show the mean LLM-coded score for partisan content salience, partisan identity salience, disagreement, and concessionary behaviour in the covert and overt conditions. Scores range from 0 to 4, with higher values indicating greater partisan content salience, partisan identity salience, disagreement, or concessionary behaviour. Error bars represent standard errors of the mean.}
\label{fig:chat_scores}
\end{figure}

Figure~\ref{fig:chat_scores} shows the mean score for each chat dimension in the covert and overt conditions. 

\subsection*{Additional Figures}

\begin{figure}[H]
  \centering
      \includegraphics[width=0.6\textwidth]{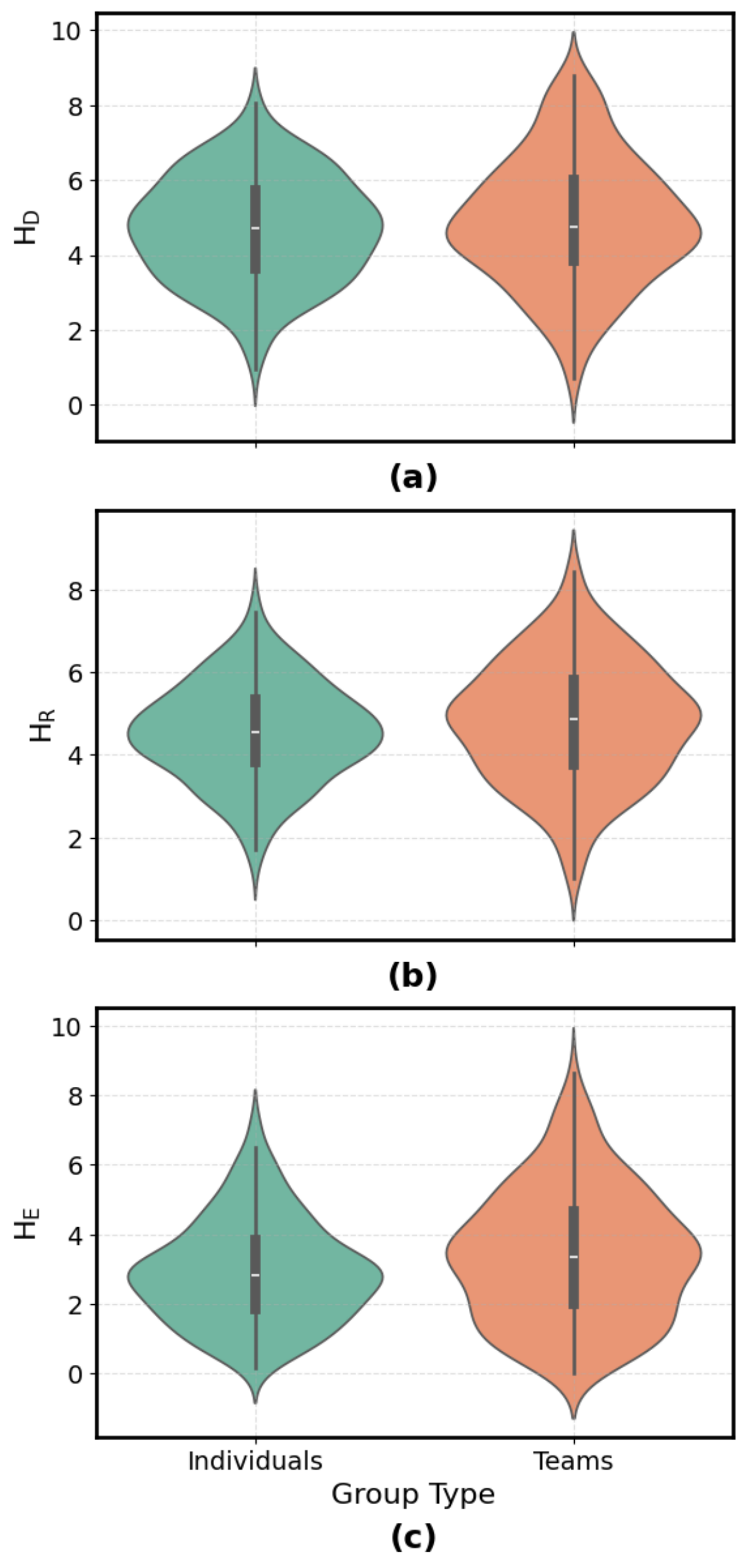}
\caption{(a) The distribution of helpfulness scores for individual versus collaborative notes based on expert evaluations. (b) The distribution of helpfulness scores for individual versus collaborative notes, based on crowd-sourced ratings from evaluators who self-identified as Democrats. (c) The distribution of  helpfulness scores for individual versus collaborative notes by tweet source and group configuration, based on crowd-sourced ratings from evaluators who self-identified as Republicans.}
\label{IndivGroupsbyTreatment}
\end{figure}

\begin{figure}[H]
  \centering
      \includegraphics[width=0.6\textwidth]{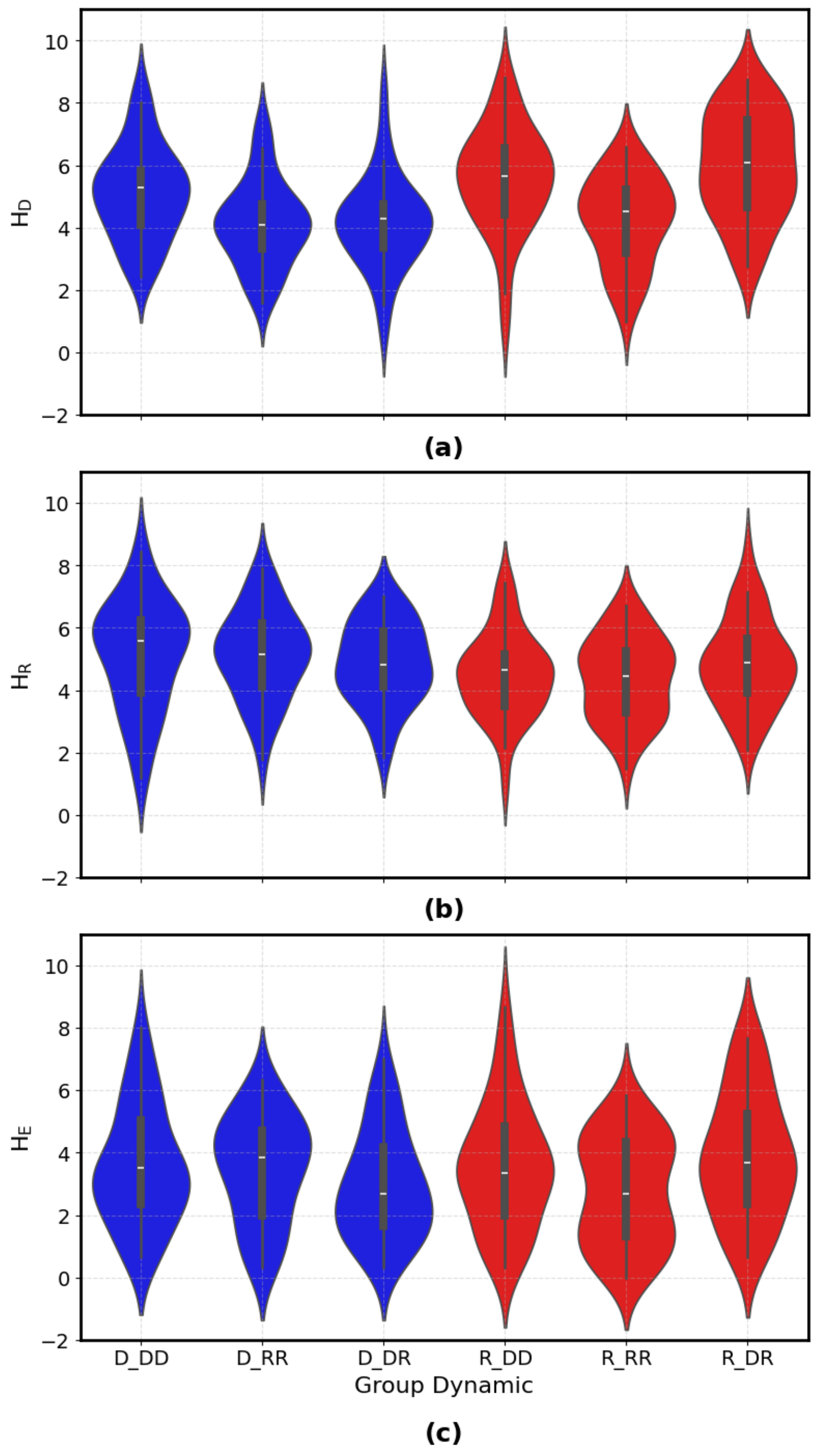}
\caption{(a) The distribution of helpfulness scores for notes across different group configurations, based on ratings from expert evaluators. (b) The distribution of helpfulness scores for notes across different group configurations, based on crowd-sourced ratings from evaluators who self-identified as Democrats. (c) The distribution of helpfulness scores for notes across different group configurations, based on crowd-sourced ratings from evaluators who self-identified as Republicans. The notation on the x-axis follows the format "post partisanship\_team partisanship." For example, if a Democrat and a Republican evaluated a Democrat post, they would be labelled as D\_DR. Similarly, if two Republicans evaluated a Republican post, they would be labelled as R\_RR.}
\label{GroupsbyTreatment}
\end{figure}


\begin{figure}[H]
  \centering
      \includegraphics[width=0.8\textwidth]{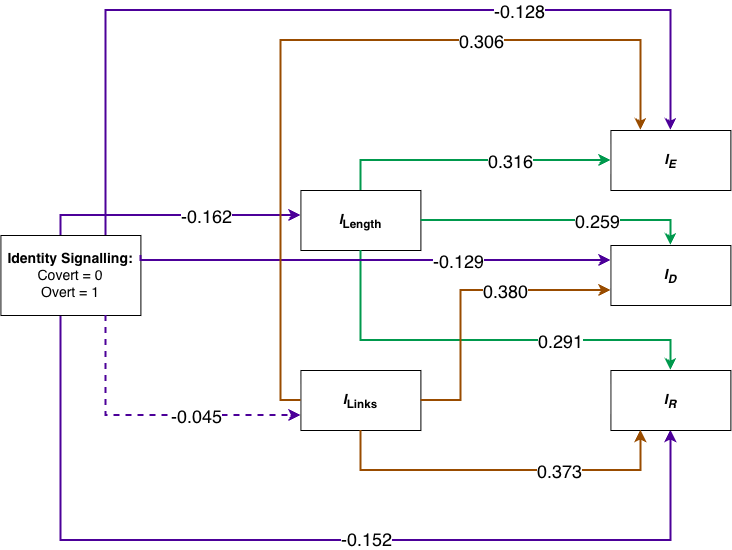}
\caption{Path diagram of political identity signalling, changes in note characteristics, and changes in note helpfulness.
Labels show standardized OLS coefficients ($\beta$). Solid arrows indicate $p < 0.05$; dashed arrows indicate $p \geq 0.05$.
Purple, green, and orange arrows represent paths from identity signalling, changes in note length, and changes in the number of links, respectively. Coefficients are estimated from separate regressions reported in Tables~S9 and~S11.}
\label{fig:identity_signalling_SEM}
\end{figure}

\subsection*{Additional Tables}
\FloatBarrier

\begin{table}[!htbp]
\centering
\caption{\textbf{Helpfulness Ratings Between Teams and Individuals} Means and standard deviations (in parentheses) of the helpfulness of individually and collaboratively written notes according to experts ($H_E$), Democrats ($H_D$), and Republicans ($H_R$).}

\label{collaboration_table}
\begin{tabular}{@{}ccccccc@{}}
\toprule
 & \textbf{Individuals (SD)} 
 & \textbf{Teams (SD)} 
 & \textbf{$t$} 
 & \textbf{Cohen's $d$}
 & \textbf{\begin{tabular}[c]{@{}c@{}}$p$\\ (parametric)\end{tabular}} 
 & \textbf{\begin{tabular}[c]{@{}c@{}}$p$\\ (non-parametric)\end{tabular}} \\ 
\midrule
$H_E$ & 2.9 (2.0) & 3.4 (1.9) & -2.8 & 0.27 & 0.0049 & 0.0018 \\
$H_D$ & 4.7 (1.9) & 4.9 (1.7) & -1.3 & 0.13 & 0.1844 & 0.0926 \\
$H_R$ & 4.6 (1.5) & 4.8 (1.5) & -2.0 & 0.19 & 0.0456 & 0.0240 \\
\bottomrule
\end{tabular}
\end{table}

\begin{table}[!htbp]
\centering
\caption{\textbf{Changes in Helpfulness Between Different Team Political Composition} Means and standard deviations (in parentheses) of ${I}_{\rm{E}}$ for teams with different political compositions: Democrat teams (DD), mixed teams (DR), and Republican teams (RR), evaluated separately for Democrat and Republican posts. 
In each comparison, the reference category is DD for Democrat posts and RR for Republican posts.}
\label{political_composition_table}
\begin{adjustbox}{width=\textwidth}
\begin{tabular}{@{}lccccccccc@{}}
\toprule
\multicolumn{1}{c}{\textbf{Post}} & \textbf{Control} & \textbf{Treatment} & \textbf{$N_c$} & \textbf{$N_t$} & \textbf{${I}_{\rm{E}}$ (SD), Control} & \textbf{${I}_{\rm{E}}$ (SD), Treatment} & \textbf{$t$} & \textbf{\begin{tabular}[c]{@{}c@{}}$p$\\ (parametric)\end{tabular}} & \textbf{\begin{tabular}[c]{@{}c@{}}$p$\\ (non-parametric)\end{tabular}} \\ \midrule
\multirow{2}{*}{Republican}        & RR               & DR                 & 35             & 37             & 0.11 (1.3)                    & 0.78 (1.5)                    & -2.0             & 0.0463                                                                  & 0.0184                                                                      \\
                                   & RR               & DD                 & 35             & 37             & 0.11 (1.3)                    & 0.32 (1.5)                    & -0.6             & 0.5206                                                                  & 0.2517                                                                      \\
\multirow{2}{*}{Democrat}          & DD               & DR                 & 36             & 35             & 0.36 (1.5)                    & 0.5 (1.8)                     & -0.3             & 0.7275                                                                  & 0.3596                                                                      \\
                                   & DD               & RR                 & 36             & 32             & 0.36 (1.5)                    & 0.86 (1.6)                    & -1.3             & 0.1845                                                                  & 0.0845                                                                      \\ \bottomrule
\end{tabular}
\end{adjustbox}
\end{table}

\begin{table}[!htbp]
\centering
\caption{\textbf{Changes in Helpfulness Between Teams in Covert and Overt Condition} Means and standard deviations (in parentheses) of $I_E$, $I_D$, $I_R$  between Covert and Overt treatments.
}
\label{covert_overt_table}
\begin{tabular}{@{}cccccc@{}}
\toprule
 & \textbf{Covert (SD)} & \textbf{Overt (SD)} & \textbf{$t$} & \textbf{\begin{tabular}[c]{@{}c@{}}$p$\\ (parametric)\end{tabular}} & \textbf{\begin{tabular}[c]{@{}c@{}}$p$\\ (non-parametric)\end{tabular}} \\ \midrule
$I_E$             & 0.76 (1.5)                                                               & 0.16 (1.5)                                                              & 2.9              & 0.0043                                                                  & 0.0021                                                                      \\
$I_D$             & 0.45 (1.3)                                                               & -0.045 (1.3)                                                            & 2.8              & 0.0055                                                                  & 0.0023                                                                     \\
$I_R$             & 0.54 (1.4)                                                               & -0.037 (1.2)                                                            & 3.2              & 0.0014                                                                 & 0.0003                                                                     \\ \bottomrule
\end{tabular}

\end{table}

\begin{table}[!htbp]
\centering
\caption{\textbf{Differences in Helpfulness Scores Between Individuals and Teams by Group Partisan Dynamics.} Statistical comparison of the helpfulness scores between individuals and teams across different groups. The mean and standard deviation (in parentheses) are presented for each distribution, along with the t-value and p-value.}
\begin{tabular}{llrrrr}
\toprule
\textbf{Group (N)} & \textbf{Variable} & \textbf{Individual} & \textbf{Teams} & \textbf{$t$} & \textbf{$p$} \\
\midrule
D\_DD (N = 36) & $H_{\rm D}$ & 5.1 (1.5) & 5.1 (1.2) & -0.078 & 0.93 \\
               & $H_{\rm R}$ & 5.1 (1.8) & 4.8 (1.2) & 0.98 & 0.33 \\
               & $H_{\rm E}$ & 3.6 (1.9) & 3.2 (1.5) & 0.89 & 0.38 \\
\midrule
D\_RR (N = 32) & $H_{\rm D}$ & 4.1 (1.4) & 4.0 (1.4) & 0.62 & 0.54 \\
               & $H_{\rm R}$ & 5.1 (1.4) & 4.9 (1.3) & 0.70 & 0.49 \\
               & $H_{\rm E}$ & 3.5 (1.7) & 2.6 (1.3) & 2.2 & 0.028 \\
\midrule
D\_RD (N = 35) & $H_{\rm D}$ & 4.2 (1.5) & 4.0 (1.2) & 0.42 & 0.68 \\
               & $H_{\rm R}$ & 4.9 (1.3) & 4.3 (1.1) & 2.0 & 0.05 \\
               & $H_{\rm E}$ & 2.9 (1.7) & 2.4 (1.1) & 1.4 & 0.16 \\
\midrule
R\_DD (N = 37) & $H_{\rm D}$ & 5.6 (1.7) & 5.1 (1.2) & 1.2 & 0.24 \\
               & $H_{\rm R}$ & 4.5 (1.4) & 4.8 (1.2) & -0.92 & 0.36 \\
               & $H_{\rm E}$ & 3.6 (1.2) & 3.2 (1.5) & 0.88 & 0.38 \\
\midrule
R\_RR (N = 37) & $H_{\rm D}$ & 4.2 (1.4) & 3.9 (1.4) & 0.98 & 0.33 \\
               & $H_{\rm R}$ & 4.4 (1.3) & 4.9 (1.3) & -1.7 & 0.095 \\
               & $H_{\rm E}$ & 2.8 (1.7) & 2.6 (1.3) & 0.36 & 0.72 \\
\midrule
R\_RD (N = 39) & $H_{\rm D}$ & 6.0 (1.7) & 4.0 (1.2) & 5.7 & $2.2 \times 10^{-7}$ \\
               & $H_{\rm R}$ & 4.9 (1.4) & 4.3 (1.1) & 2.0 & 0.050 \\
               & $H_{\rm E}$ & 3.9 (2.0) & 2.4 (1.1) & 3.7 & 0.00035 \\
\bottomrule
\end{tabular}
\label{tab:stats_ind_vs_groups}
\end{table}

\begin{table}[!htbp]
    \centering
        \caption{\textbf{Effects of Team Composition and Post Political Partisanship on $I_{\rm D}$, $I_{\rm R}$, and $I_{\rm E}$.} Results of the $t$-tests for $I_{\rm D}$, $I_{\rm R}$ and $I_{\rm E}$ for different team compositions and posts political partisanships. The mean and standard deviation (in parentheses) are presented for each distribution, along with the $t$-value and $p$-value.}
    \begin{tabular}{lrrrrrr}
        \hline
        Pair & $t$ ($I_{\rm D}$) & $p$ ($I_{\rm D}$) & $t$ ($I_{\rm R}$) & $p$ ($I_{\rm R}$) & $t$ ($I_{\rm E}$) & $p$ ($I_{\rm E}$) \\
        \hline
        D\_DD vs D\_RR & -0.837 & 0.406 &  0.285 & 0.777 & -1.341 & 0.185 \\
        D\_DD vs D\_DR & -0.520 & 0.605 & -0.742 & 0.461 & -0.350 & 0.727 \\
        D\_DD vs R\_DD & -0.400 & 0.690 &  0.430 & 0.668 &  0.092 & 0.927 \\
        D\_DD vs R\_RR &  0.228 & 0.820 &  1.033 & 0.305 &  1.000 & 0.321 \\
        D\_DD vs R\_DR & -2.876 & 0.005 &  0.606 & 0.547 & -1.055 & 0.295 \\
        D\_RR vs D\_DR &  0.253 & 0.801 & -1.064 & 0.291 &  0.877 & 0.384 \\
        D\_RR vs R\_DD &  0.394 & 0.695 &  0.095 & 0.924 &  1.455 & 0.150 \\
        D\_RR vs R\_RR &  1.065 & 0.291 &  0.719 & 0.475 &  2.395 & 0.019 \\
        D\_RR vs R\_DR & -2.023 & 0.047 &  0.284 & 0.777 &  0.371 & 0.712 \\
        D\_DR vs R\_DD &  0.125 & 0.901 &  1.397 & 0.167 &  0.439 & 0.662 \\
        D\_DR vs R\_RR &  0.732 & 0.466 &  2.006 & 0.049 &  1.264 & 0.210 \\
        D\_DR vs R\_DR & -2.169 & 0.033 &  1.527 & 0.131 & -0.596 & 0.553 \\
        R\_DD vs R\_RR &  0.618 & 0.538 &  0.760 & 0.449 &  0.922 & 0.359 \\
        R\_DD vs R\_DR & -2.364 & 0.021 &  0.233 & 0.816 & -1.170 & 0.246 \\
        R\_RR vs R\_DR & -3.112 & 0.003 & -0.496 & 0.621 & -2.154 & 0.035 \\
        \hline
    \end{tabular}

    \label{tab:t_tests}
\end{table}

\begin{table}[!htbp]
\centering
\caption{\textbf{Changes in Note Helpfulness Between the Covert and Overt Conditions.} Comparison of changes in the helpfulness of notes for notes written in the covert and overt treatment. The mean and standard deviation (in parentheses) are presented for each distribution, along with the $t$-value and $p$-value.}
\begin{tabular}{llrrrr}
\toprule
\textbf{Group} & \textbf{Variable} & \textbf{covert} & \textbf{Overt} & \textbf{$t$} & \textbf{$p$} \\
\midrule
D\_DD (N = 36) & $I_{\rm D}$ & 0.061 (1.0) & -0.11 (1.4) & 0.43 & 0.67 \\
               & $I_{\rm R}$ & 1.0 (1.5) & -0.34 (1.4) & 2.8 & 0.0079 \\
               & $I_{\rm E}$ & 0.88 (1.5) & -0.17 (1.3) & 2.2 & 0.036 \\
\midrule
D\_RR (N = 32) & $I_{\rm D}$ & 0.33 (1.3) & 0.12 (1.1) & 0.48 & 0.64 \\
               & $I_{\rm R}$ & 0.39 (1.9) & 0.12 (1.2) & 0.46 & 0.65 \\
               & $I_{\rm E}$ & 0.98 (1.6) & 0.78 (1.6) & 0.34 & 0.73 \\
\midrule
D\_RD (N = 39) & $I_{\rm D}$ & 0.53 (1.6) & -0.24 (1.1) & 1.6 & 0.11 \\
               & $I_{\rm R}$ & 0.90 (1.4) & 0.32 (1.1) & 1.4 & 0.18 \\
               & $I_{\rm E}$ & 0.93 (2.2) & 0.083 (1.3) & 1.4 & 0.18 \\
\midrule
R\_DD (N = 37) & $I_{\rm D}$ & 0.24 (1.0) & -0.062 (1.6) & 0.67 & 0.51 \\
               & $I_{\rm R}$ & 0.34 (0.90) & 0.074 (1.3) & 0.71 & 0.48 \\
               & $I_{\rm E}$ & 0.75 (1.3) & -0.13 (1.5) & 1.9 & 0.069 \\
\midrule
R\_RR (N = 37) & $I_{\rm D}$ & 0.33 (1.3) & -0.53 (0.99) & 2.3 & 0.023 \\
               & $I_{\rm R}$ & 0.35 (1.3) & -0.38 (1.1) & 1.8 & 0.077 \\
               & $I_{\rm E}$ & 0.079 (1.3) & -0.028 (1.34) & 0.25 & 0.81 \\
\midrule
R\_RD (N = 39) & $I_{\rm D}$ & 1.1 (1.4) & 0.55 (1.4) & 1.2 & 0.23 \\
               & $I_{\rm R}$ & 0.29 (1.4) & -0.022 (1.2) & 0.75 & 0.46 \\
               & $I_{\rm E}$ & 0.98 (1.3) & 0.42 (1.8) & 1.1 & 0.27 \\
\bottomrule
\end{tabular}
\label{tab:stats_cov_vs_ov}
\end{table}

\begin{table}[!htbp]
\centering
\caption{\textbf{Regression Results of Changes in Note Helpfulness as a Function of Note Characteristics Metrics.} Regression results for the changes in the metrics evaluated by Democrats, Republicans, and experts. The independent variables include the change in the length of the note ($I_{\rm{Length}}$), the change in the number of links used in the note ($I_{\rm{Links}}$), the change in the sophistication level of the note ($I_{\rm{Level}}$), and the change in use of numbers in the note ($I_{\rm{Stats}}$). The values in the parentheses are the standard errors. The Adjusted $\rm R^2$ and F-statistics of each dependent variable are listed.}
\begin{adjustbox}{width=\textwidth}
\begin{tabular}{ccccccc}
\hline
\multirow{2}{*}{Ordinary Least Squares (OLS) Regression} 
& \multicolumn{2}{c}{$I_{\rm E}$} 
& \multicolumn{2}{c}{$I_{\rm D}$} 
& \multicolumn{2}{c}{$I_{\rm R}$} \\ 
\cline{2-7}
& Coefficient & $p$ 
& Coefficient & $p$ 
& Coefficient & $p$ \\ 
\hline
$I_{\rm{Links}}$  
& 1.071 (0.226) & \textless{}0.001 
& 1.197 (0.192) & \textless{}0.001 
& 1.208 (0.195) & \textless{}0.001 \\ 
\hline
$I_{\rm{Length}}$ 
& 0.031 (0.006) & \textless{}0.001 
& 0.023 (0.005) & \textless{}0.001 
& 0.027 (0.005) & \textless{}0.001 \\ 
\hline
$I_{\rm{Stats}}$ 
& 0.582 (0.317) & 0.068 
& 0.001 (0.269) & 0.998 
& 0.214 (0.273) & 0.435 \\ 
\hline
$I_{\rm{Level}}$ 
& 0.047 (0.029) & 0.105 
& 0.033 (0.024) & 0.173 
& -0.009 (0.025) & 0.716 \\ 
\hline
Constant 
& 0.401 (0.096) & \textless{}0.001 
& 0.143 (0.081) & 0.079 
& 0.229 (0.082) & 0.006 \\ 
\hline
Observations 
& \multicolumn{2}{c}{216} 
& \multicolumn{2}{c}{216} 
& \multicolumn{2}{c}{216} \\ 
\hline
$R^2$ 
& \multicolumn{2}{c}{0.239} 
& \multicolumn{2}{c}{0.238} 
& \multicolumn{2}{c}{0.251} \\ 
\hline
Adjusted $R^2$ 
& \multicolumn{2}{c}{0.225} 
& \multicolumn{2}{c}{0.224} 
& \multicolumn{2}{c}{0.237} \\ 
\hline
Residual Std. Error 
& \multicolumn{2}{c}{1.363 (df = 211)} 
& \multicolumn{2}{c}{1.157 (df = 211)} 
& \multicolumn{2}{c}{1.174 (df = 211)} \\ 
\hline
F Statistic 
& \multicolumn{2}{c}{16.580 (df = 4; 211)} 
& \multicolumn{2}{c}{16.510 (df = 4; 211)} 
& \multicolumn{2}{c}{17.710 (df = 4; 211)} \\ 
\hline
\end{tabular}
\end{adjustbox}
\label{tab:quant_regression}
\end{table}

\begin{table}[!htbp]
\centering
\caption{\textbf{Regression Results of Changes in Note Helpfulness as a Function of Post Alignment.} Regression results for the changes in the metrics evaluated by Democrats, Republicans, and experts as a function of post alignment. Post alignment is modelled as an ordered variable with three levels (0,1,2). The values in the parentheses are the standard errors. The table reports the linear and quadratic trend coefficients from polynomial contrasts, along with the Adjusted $\rm R^2$ and F-statistic of the model.}
\label{tab:reg_post_alignment}
\begin{adjustbox}{width=\textwidth}
\begin{tabular}{ccccccc}
\hline
\multirow{2}{*}{Ordinary Least Squares (OLS) Regression} 
& \multicolumn{2}{c}{$I_{\rm E}$} 
& \multicolumn{2}{c}{$I_{\rm D}$} 
& \multicolumn{2}{c}{$I_{\rm R}$} \\ 
\cline{2-7}
& Coefficient & $p$ 
& Coefficient & $p$ 
& Coefficient & $p$ \\ 
\hline
Post Alignment (Linear) 
& -0.273 (0.183) & 0.138 
& -0.147 (0.154) & 0.340 
& -0.037 (0.160) & 0.816 \\ 
\hline
Post Alignment (Quadratic) 
& -0.192 (0.181) & 0.290 
& -0.375 (0.152) & 0.015 
& -0.134 (0.158) & 0.397 \\ 
\hline
Constant 
& 0.461 (0.105) & \textless{}0.001 
& 0.199 (0.088) & 0.025 
& 0.251 (0.092) & 0.007 \\ 
\hline
Observations 
& \multicolumn{2}{c}{216} 
& \multicolumn{2}{c}{216} 
& \multicolumn{2}{c}{216} \\ 
\hline
$R^2$ 
& \multicolumn{2}{c}{0.016} 
& \multicolumn{2}{c}{0.032} 
& \multicolumn{2}{c}{0.004} \\ 
\hline
Adjusted $R^2$ 
& \multicolumn{2}{c}{0.006} 
& \multicolumn{2}{c}{0.023} 
& \multicolumn{2}{c}{-0.006} \\ 
\hline
Residual Std. Error 
& \multicolumn{2}{c}{1.543 (df = 213)} 
& \multicolumn{2}{c}{1.299 (df = 213)} 
& \multicolumn{2}{c}{1.348 (df = 213)} \\ 
\hline
F Statistic 
& \multicolumn{2}{c}{1.696 (df = 2; 213)} 
& \multicolumn{2}{c}{3.531 (df = 2; 213)} 
& \multicolumn{2}{c}{0.391 (df = 2; 213)} \\ 
\hline
\end{tabular}
\end{adjustbox}
\end{table}

\begin{table}[!htbp]
\centering
\caption{\textbf{Regression Results of Changes in Note Characteristics
as a Function of Political Identity Signalling.}
Separate ordinary least squares (OLS) regressions predicting changes
in note length and number of links from political identity signalling,
with covert signalling as the reference category.
Standard errors are reported in parentheses.
$p$-values are based on two-sided tests.}
\label{tab:identity_note_characteristics}

\begin{tabular}{lcccc}
\hline
\multirow{2}{*}{Ordinary Least Squares (OLS) Regression} 
& \multicolumn{2}{c}{$I_{\mathrm{Length}}$} 
& \multicolumn{2}{c}{$I_{\mathrm{Links}}$}  \\ 
\cline{2-5}
& Coefficient & $p$ 
& Coefficient & $p$ \\ \hline
Identity signalling
& -5.032 (2.090) & 0.0169
& -0.037 (0.057) & 0.514 \\ \hline

Constant
& 1.991 (1.478) & 0.179
& 0.056 (0.040) & 0.167 \\ 

\midrule
Observations
& \multicolumn{2}{c}{216}
& \multicolumn{2}{c}{216} \\ \hline

$R^2$
& \multicolumn{2}{c}{0.026}
& \multicolumn{2}{c}{0.002} \\ \hline

Adjusted $R^2$
& \multicolumn{2}{c}{0.022}
& \multicolumn{2}{c}{ -0.003} \\ \hline

Residual Std.\ Error
& \multicolumn{2}{c}{15.360 (df= 214)}
& \multicolumn{2}{c}{0.427 (df = 214)} \\ \hline

$F$-statistic
& \multicolumn{2}{c}{5.798 (df = 1; 214)}
& \multicolumn{2}{c}{0.427 (df = 1; 214)} \\

\bottomrule
\end{tabular}
\end{table}

\begin{table}[!htbp]
\centering
\caption{\textbf{Regression Results of Changes in Note Helpfulness as a Function of Political Signalling.} Regression results for the changes in the metrics evaluated by Democrats, Republicans, and experts. The independent variable is the indicator for identity signalling (covert as the reference category). The values in the parentheses are the standard errors. The adjusted $\rm R^2$ and F-statistics for each dependent variable are listed.}
\begin{adjustbox}{width=\textwidth}
\begin{tabular}{ccccccc}
\hline
\multirow{2}{*}{Ordinary Least Squares (OLS) Regression} 
& \multicolumn{2}{c}{$I_{\rm E}$} 
& \multicolumn{2}{c}{$I_{\rm D}$} 
& \multicolumn{2}{c}{$I_{\rm R}$} \\ 
\cline{2-7}
& Coefficient & $p$ 
& Coefficient & $p$ 
& Coefficient & $p$ \\ 
\hline
Identity Signalling
& -0.597 (0.207) & 0.004
& -0.494 (0.176) & 0.005
& -0.579 (0.179) & 0.001 \\ 
\hline
Constant 
& 0.758 (0.146) & \textless{}0.001 
& 0.448 (0.124) &  \textless{}0.001 
& 0.542 (0.127) & \textless{}0.001 \\ 
\hline
Observations 
& \multicolumn{2}{c}{216} 
& \multicolumn{2}{c}{216} 
& \multicolumn{2}{c}{216} \\ 
\hline
$R^2$ 
& \multicolumn{2}{c}{0.037} 
& \multicolumn{2}{c}{0.035} 
& \multicolumn{2}{c}{0.047} \\ 
\hline
Adjusted $R^2$ 
& \multicolumn{2}{c}{0.033} 
& \multicolumn{2}{c}{0.031} 
& \multicolumn{2}{c}{0.042} \\ 
\hline
Residual Std. Error 
& \multicolumn{2}{c}{1.522 (df = 214)} 
& \multicolumn{2}{c}{1.293 (df = 214)} 
& \multicolumn{2}{c}{1.315 (df = 214)} \\ 
\hline
F Statistic 
& \multicolumn{2}{c}{8.31 (df = 1; 214)} 
& \multicolumn{2}{c}{7.874 (df = 1; 214)} 
& \multicolumn{2}{c}{10.48 (df = 1; 214)} \\ 
\hline
\end{tabular}
\end{adjustbox}
\end{table}

\begin{table}[!htbp]
\centering
\caption{\textbf{Regression Results of Changes in Note Helpfulness as a Function of Political Signalling and Note Characteristics.} Regression results for the changes in the metrics evaluated by Democrats, Republicans, and experts. The independent variables include an indicator for identity signalling (covert as the reference category), the change in the length of the note ($I_{\rm{Length}}$), and the change in the number of links used in the note ($I_{\rm{Links}}$). The values in the parentheses are the standard errors. The adjusted $\rm R^2$ and F-statistics for each dependent variable are listed.}
\begin{adjustbox}{width=\textwidth}
\begin{tabular}{ccccccc}
\hline
\multirow{2}{*}{Ordinary Least Squares (OLS) Regression} 
& \multicolumn{2}{c}{$I_{\rm E}$} 
& \multicolumn{2}{c}{$I_{\rm D}$} 
& \multicolumn{2}{c}{$I_{\rm R}$} \\ 
\cline{2-7}
& Coefficient & $p$ 
& Coefficient & $p$ 
& Coefficient & $p$ \\ 
\hline
Identity Signalling
& -0.397 (0.189) & 0.037
& -0.339 (0.158) & 0.033
& -0.408 (0.159) & 0.011 \\ 
\hline
$I_{\rm{Links}}$  
& 1.138 (0.224) & \textless{}0.001 
& 1.199 (0.188) & \textless{}0.001 
& 1.205 (0.190) & \textless{}0.001 \\ 
\hline
$I_{\rm{Length}}$ 
& 0.031 (0.006) & \textless{}0.001 
& 0.022 (0.005) & \textless{}0.001 
& 0.025 (0.005) & \textless{}0.001 \\ 
\hline
Constant 
& 0.632 (0.133) & \textless{}0.001 
& 0.338 (0.111) & 0.003
& 0.425 (0.112) & \textless{}0.001 \\ 
\hline
Observations 
& \multicolumn{2}{c}{216} 
& \multicolumn{2}{c}{216} 
& \multicolumn{2}{c}{216} \\ 
\hline
$R^2$ 
& \multicolumn{2}{c}{0.231} 
& \multicolumn{2}{c}{0.248} 
& \multicolumn{2}{c}{0.271} \\ 
\hline
Adjusted $R^2$ 
& \multicolumn{2}{c}{0.220} 
& \multicolumn{2}{c}{0.237} 
& \multicolumn{2}{c}{0.261} \\ 
\hline
Residual Std. Error 
& \multicolumn{2}{c}{1.367 (df = 212)} 
& \multicolumn{2}{c}{1.147 (df = 212)} 
& \multicolumn{2}{c}{1.155 (df = 212)} \\ 
\hline
F Statistic 
& \multicolumn{2}{c}{21.210 (df = 3; 212)} 
& \multicolumn{2}{c}{23.280 (df = 3; 212)} 
& \multicolumn{2}{c}{26.320 (df = 3; 212)} \\ 
\hline
\end{tabular}
\end{adjustbox}
\end{table}

\begin{table}[!htbp]
\centering
\caption{\textbf{Regression Results of Changes in Note Helpfulness as a Function of Political Signalling and Political Diversity.} Regression results for changes in note helpfulness as evaluated by Democrats, Republicans, and experts. The models include interaction terms between political identity signalling and team diversity. The reference category is overt identity signalling among politically aligned teams. The values in the parentheses are the standard errors. The adjusted $\rm R^2$ and F-statistics for each dependent variable are listed.}
\label{tab:ols_helpfulness_improvement_session_diversity}
\begin{adjustbox}{width=\textwidth}
\begin{tabular}{ccccccc}
\hline
\multirow{2}{*}{Ordinary Least Squares (OLS) Regression} 
& \multicolumn{2}{c}{$I_{\rm E}$} 
& \multicolumn{2}{c}{$I_{\rm D}$} 
& \multicolumn{2}{c}{$I_{\rm R}$} \\ 
\cline{2-7}
& Coefficient & $p$ 
& Coefficient & $p$ 
& Coefficient & $p$ \\ 
\hline
Covert $\times$ Diverse
& 0.847 (0.306) & 0.006
& 0.983 (0.257) & \textless{}0.001
& 0.692 (0.264) & 0.009 \\ 
\hline
Overt $\times$ Diverse
& 0.141 (0.311) & 0.651
& 0.299 (0.261) & 0.254
& 0.278 (0.269) & 0.303 \\ 
\hline
Covert $\times$ Aligned
& 0.534 (0.256) & 0.038
& 0.382 (0.215) & 0.077
& 0.661 (0.221) & 0.003 \\ 
\hline
Constant
& 0.113 (0.180) & 0.529
& -0.145 (0.151) & 0.337
& -0.130 (0.155) & 0.403 \\ 
\hline
Observations 
& \multicolumn{2}{c}{216} 
& \multicolumn{2}{c}{216} 
& \multicolumn{2}{c}{216} \\ 
\hline
$R^2$ 
& \multicolumn{2}{c}{0.043} 
& \multicolumn{2}{c}{0.065} 
& \multicolumn{2}{c}{0.052} \\ 
\hline
Adjusted $R^2$ 
& \multicolumn{2}{c}{0.029} 
& \multicolumn{2}{c}{0.052} 
& \multicolumn{2}{c}{0.038} \\ 
\hline
Residual Std. Error 
& \multicolumn{2}{c}{1.525 (df = 212)} 
& \multicolumn{2}{c}{1.279 (df = 212)} 
& \multicolumn{2}{c}{1.318 (df = 212)} \\ 
\hline
F Statistic 
& \multicolumn{2}{c}{3.174 (df = 3; 212)} 
& \multicolumn{2}{c}{4.929 (df = 3; 212)} 
& \multicolumn{2}{c}{3.837 (df = 3; 212)} \\ 
\hline
\end{tabular}
\end{adjustbox}
\end{table}
\end{document}